\documentclass[fleqn,usenatbib,nofootinbib, twocolumn]{mnras}
\usepackage{graphicx}
\usepackage{amsmath,amssymb}
\usepackage{lastpage}
\usepackage[all]{hypcap}
\usepackage[T1]{fontenc}
\usepackage{float}
\usepackage{subfigure}
\usepackage{afterpage}
\usepackage[usenames]{color}
\usepackage{mathrsfs}
\usepackage{upgreek}
\usepackage{hyperref}
\usepackage{gensymb}
\usepackage{color}
\usepackage[dvipsnames]{xcolor}
\usepackage{longtable}
\usepackage{threeparttable}
\usepackage{multirow}
\usepackage{placeins}
\usepackage{bm}
\usepackage[capitalise]{cleveref}
\usepackage{enumitem}
\usepackage{mathtools}

\usepackage{tabularx}

\gdef\rowstyle{}
\newcommand{\gmark}{\gdef\rowstyle{\color{gray}}}
\newcommand{\nmark}{\gdef\rowstyle{}}

\newcommand{\oxford}{Astrophysics, University of Oxford, DWB, Keble Road, Oxford OX1 3RH, United Kingdom}
\newcommand{\iap}{CNRS \& Sorbonne Universit\'e, Institut d’Astrophysique de Paris (IAP), UMR 7095, 98 bis bd Arago, F-75014 Paris, France}
\newcommand{\icg}{Institute of Cosmology \& Gravitation, University of Portsmouth, Portsmouth PO1 3FX, United Kingdom}
\newcommand{\soton}{Optoelectronic Research Centre, University of Southampton, Southampton SO17 1BJ, United Kingdom}

\title[Functional form of mass/luminosity functions]{The functional form of galaxy and halo luminosity and mass functions}
\author[Ford, Desmond, Bartlett \& Ferreira]{
Amelia~Ford$^{1,2}$, Harry~Desmond$^{1}$\thanks{harry.desmond@port.ac.uk}, Deaglan~J.~Bartlett$^{3,4}$ and Pedro~G.~Ferreira$^4$\\
$^{1}$\icg\\
$^{2}$\soton\\
$^{3}$\iap\\
$^{4}$\oxford\\
}

\pubyear{2026}

\begin{document}
\label{FirstPage}
\pagerange{\pageref{FirstPage}--\pageref{LastPage}}
\maketitle

\begin{abstract}
The galaxy luminosity and stellar mass function (LF, SMF), and halo mass function (HMF), are fundamental quantities in astrophysics and crucial inputs to a range of astrophysical and cosmological analyses. They are typically parametrised by fitting functions that have been chosen ``by eye'' to match observed or simulated data. We apply symbolic regression---specifically the Exhaustive Symbolic Regression (ESR) algorithm---to automate the search for optimal LF, SMF and HMF functional forms. ESR scores all functions up to a maximum complexity composed of a user-defined basis set of operators using the description length, an approximation to the Bayesian evidence that balances accuracy with complexity. We find many functions outperforming the Schechter and double Schechter functions for the LF and SMF, and that outperform the Press--Schechter and Warren/Tinker functions for the HMF. By additionally imposing ``physicality checks'' on functions' extrapolation and integration properties, we identify the optimal, low-complexity functional forms in terms of accuracy, simplicity and behaviour beyond the data range. As well as providing drop-in replacements for literature LF, SMF and HMF fitting functions, and identifying robust behaviour across well-fitting functions, we present a framework with which symbolic regression may be used to automate the discovery of optimal functions for any astrophysical dataset.
\end{abstract}


\section{Introduction}
\label{sec:intro}

Structures in the Universe are usefully described by their $N$-point statistics, characterising their abundance and clustering. The $1$-point statistic gives the number of objects per unit mass or luminosity per unit volume, and is called a mass or luminosity function depending on the quantity the abundance is conditioned on. These are fundamental quantities in cosmology~\citep{1980lssu.book.....P}.

The \emph{halo mass function (HMF)} describes the number density of dark matter haloes as a function of halo mass. When small perturbations seeded by inflation grow sufficiently massive through gravitational accretion, they decouple from the expansion of the Universe and collapse to form bound, virialised haloes. The HMF may therefore be predicted from the statistics of density perturbations that exceed the threshold for collapse at a given time, as described by an excursion set formalism \citep{2002PhR...372....1C}. However, this involves analytic approximations, such as spherical top-hat collapse, that introduce errors in the predicted HMF relative to that produced by the full nonlinear evolution of haloes in a cosmological context, as captured by $N$-body simulations of structure formation. This has necessitated the use of fitting formulae for cosmological analyses, as described further below.
The HMF cannot be directly inferred observationally, although there are attempts to do so using e.g. galaxy cluster abundances calibrated by X-ray or Sunyaev--Zel'dovich observations \citep{Vikhlinin_2009} and HI galaxy kinematics~\citep{Li_HMF}.

The galaxy \emph{luminosity function (LF)}, describing the abundance of galaxies as a function of their luminosity in some band, may be directly inferred from photometric galaxy surveys. First, the flux in a band is converted to luminosity through knowledge of the distance to the galaxy and any extinction by intervening matter. For sufficiently distant galaxies, distance may be determined from redshift through Hubble's law, but for nearer galaxies distance must either be measured using a redshift-independent probe such as Cepheids, Tip-of-the-Red-Giant-Branch or Type Ia supernovae, or the peculiar velocities must be corrected for using
a model for the large-scale structure of the local Universe. Combined with a model for light absorption and reddening, mostly by dust within the Milky Way, this produces a \emph{K-correction} \citep{Blanton_Roweis_2007}, the key quantity for relating flux and luminosity (or apparent and absolute magnitude). Since surveys are typically flux-limited, constructing the luminosity function requires a further correction to account for the fact that brighter galaxies can be seen to larger distance than fainter ones. This is typically achieved by calculating the volume within which one would expect to be able to see a galaxy of a certain luminosity given the flux limit of the survey, and then weighting by the inverse of this when summing galaxy abundances in bins of luminosity to calculate the luminosity function (the so-called $1/V_\text{max}$ method; \citealt{Schmidt_1968}). A more principled Bayesian alternative is described in~\citet{Kelly_2008}.

Another crucial quantity is the \emph{stellar mass function (SMF)} describing galaxy abundance as a function of stellar mass. This is more directly useful for galaxy formation studies than the LF because models first predict galaxies' stellar populations and require an additional ``mock observation'' step to convert these to luminosity predictions as a function of wavelength. Observationally, the SMF is estimated either by applying a mass-to-light ratio prescription to the galaxies' photometry, or applying a full stellar population synthesis code if (parts of) the full spectral energy distribution are available. In redder bands, the first method becomes fairly reliable \citep{Bell_2003}, although typically usage of galaxy spectra improves both the precision and accuracy of the result. The SMF is a central quantity in studies of the galaxy--halo connection \citep[e.g.][]{Behroozi_2013, Moster_2013, Wechsler_Tinker_2018} and semi-analytic modelling of galaxy formation~\citep[e.g.][]{Bower_2006, Croton_2006, Henriques_2015}, where it is typically used to constrain the models' free parameters.
Similarly, hydrodynamical simulations are unable to predict the SMF a priori, but use it as a means of tuning their sub-grid recipes and feedback schemes \citep[e.g.][]{Schaye_2015, Pillepich_2018}.
Other useful quantities include the HI mass function \citep[e.g.][]{Zwaan_2005,Anastasia_HIMF} and baryonic mass function \citep[e.g.][]{Papastergis_BMF,Eckert_BMF}, describing the abundance of galaxies as a function of neutral, HI-emitting hydrogen gas mass and total cold baryonic mass, respectively. Although amenable to treatments exactly analogous to ours, we do not consider these further here.

Use of an LF, SMF or HMF in an astrophysical or cosmological study is greatly facilitated by analytic formulae.\footnote{\citet{Kelly_2008} propose instead a non-parametric Gaussian mixture model. While this may be more accurate than simple parametric forms, it is less wieldy and interpretable.} These summarise the data in analytic functions that are simple, interpretable, and rapid to evaluate. Such functions cannot be precisely predicted a priori, so must be determined empirically from the observations or simulations. In the past, this has been done by visual inspection, guessing at functional forms likely to fit the data well with trial-and-error iteration used to identify a local optimum in the functional form. This has produced the standard fitting functions in use throughout the field: the Schechter function for the LF and SMF~\citep{Schechter} (as well as more complex alternatives such as that of~\citealt{Bernardi}), and the Warren~\citep{Warren_HMF} and reparametrised Tinker~\citep{Tinker_HMF} fits to the HMF. The original Press--Schechter formula~\citep{Press-Schechter} provides a parameter-free theoretical prediction from first principles but is inaccurate at both the high- and low-mass ends.

A more principled solution is afforded by \emph{symbolic regression (SR)}, the data-driven search for analytic expressions using machine learning. There are a few principal approaches to SR, including genetic programming, reinforcement learning and Markov Chain Monte Carlo methods (for a book-length review see~\citealt{Kronberger2024book}). SR typically aims to optimally trade-off functions' accuracies (which we define as the likelihood of the data given the function) and complexities (which we define as the number of operators, variables or parameters in it). The aim of this paper is to use symbolic regression to discover functional forms for the LF, SMF and HMF that are superior to those of the literature standards.

Since the complexity of existing fitting formulae for the LF, SMF and HMF is low, we are not seeking particularly complex functions and hence an \emph{exhaustive} search is feasible, affording a complete exploration of all functions of low complexity. We therefore use the open-source algorithm \emph{Exhaustive Symbolic Regression} (ESR;~\citealt{ESR,Desmond_ESR})\footnote{\url{https://github.com/DeaglanBartlett/ESR}} to generate and evaluate all possible simple functional fits to HMFs, LFs and SMFs. The evaluation is done using the \emph{Minimum Description Length} (MDL) principle \citep{RISSANEN1978}, which prefers functions requiring fewer bits of information to communicate the function and lower residuals of the data around the function's expectation. This allows us to find the best fitting formulae for each luminosity or mass function in a fully automated way, including a full ranking of all possible formulae by a principled metric optimally weighting accuracy and complexity.

To preview our principal findings: the best ESR function for the luminosity function (depending somewhat on the photometric pipeline used) is
\begin{equation}
	\phi_\text{LF}(x) = e^{\alpha\left(\ln\!\left(\beta + \gamma^{\,x}\right)\right)^{\!\delta}}
\end{equation}
and for the stellar mass function is
\begin{equation}
	\phi_\text{SMF}(x) = \left(\alpha + e^{\beta x^{\gamma}}\right)^{\!\delta},
\end{equation}
where $x \equiv L/10^9\,L_\odot$ or $M_\star/10^9\,M_\odot$ and $\alpha, \beta, \gamma$ and $\delta$ are fitted constants (independent between datasets). For the halo mass function (in terms of the mass variance $\sigma$ and for Friends-of-Friends haloes~\citep{FOF}) the overall best is
\begin{equation}
	f(\sigma) = \alpha/(\beta + e^{\sigma^{\gamma + \sigma}})
\end{equation}
while the best retaining the Press--Schechter-like (PS-like) $1/\sigma$ dropoff at large $\sigma$ is
\begin{equation}
	f(\sigma) = \alpha\,e^{-\beta\sigma^\gamma}\!/\sigma.
\end{equation}
Each of these outperforms the corresponding standard literature fit according to the description length criterion while displaying well-behaved extrapolation behaviour.

The structure of the paper is as follows.
In Sec.~\ref{sec:data}, we describe the observational data that we use to measure the LF and SMF, and the simulated data used to construct the HMF. In Sec.~\ref{sec:method}, we lay out our method to search systematically for the best functional forms to fit the data. Sec.~\ref{sec:results} presents the results, Sec.~\ref{sec:disc} discusses the broader ramifications, caveats and prospects of our findings and Sec.~\ref{sec:conc} concludes. We work at redshift $0$ throughout, and $\log$ has base 10.

\section{Observed and simulated data}
\label{sec:data}

\subsection{Galaxy luminosity and stellar mass functions}
\label{sec:LF_SMF}

We take luminosity data and mass-to-light ratios from~\citet{Bernardi}. This classic reference compares several different models for galaxies' surface brightness profiles using data from the Sloan Digital Sky Survey (SDSS; \citealt{York2000}), including an improved sky subtraction algorithm optimised for low-redshift galaxies~\citep{Blanton2011}. These improved fits return more light than previous approaches, boosting the LF and SMF relative to earlier results and reducing the steepness of the bright-end cutoff. The stellar mass determinations are consistent with those in the NASA-Sloan Atlas \citep{Blanton2011}. We use the $r$-band, which traces stellar mass $M_\star$ relatively well \citep{Bell_2003}.

\begin{figure}
\includegraphics[width=\columnwidth]{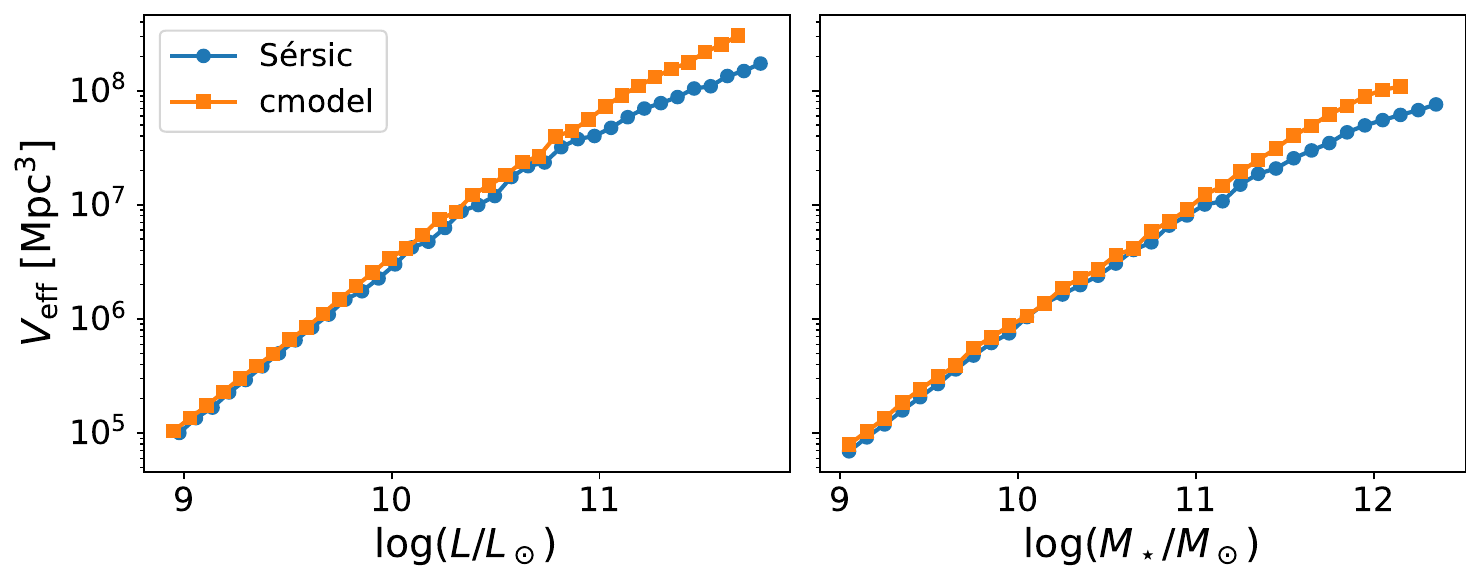}
\caption{Effective survey volume $V_\text{eff}$ for the SDSS data of~\citet{Bernardi} as a function of luminosity (left) and stellar mass (right).}
\label{fig:Veff}
\end{figure}

We consider both the single-S\'ersic (hereafter ``S\'ersic'') fit to the light profile and the ``cmodel'' photometry (an SDSS composite model combining exponential and de Vaucouleurs profiles in a linear weighting) from~\citet{Bernardi} to assess the difference in the functional form of the LF that this induces.
We use the Poisson uncertainties listed in the online tables to determine the number counts of galaxies in each bin, which constitute the target of our SR analysis. This also determines the effective volume within each bin, used to translate between counts and
the LF and SMF.
Fig.~\ref{fig:Veff} shows the effective volume $V_\text{eff}$ as a function of luminosity and stellar mass for our datasets. $V_\text{eff}$ increases steeply with luminosity and mass because brighter and more massive galaxies can be detected to larger distances.

\subsection{Halo mass functions}
\label{sec:HMF}

We take our HMFs from the \texttt{Quijote} suite of $N$-body simulations~\citep{Quijote}. Each of these is run using a $\Lambda$CDM cosmology with parameters $\Omega_{\rm m} = 0.3175$, $\Omega_{\rm \Lambda} = 0.6825$, $\Omega_{\rm b} = 0.049$, $H_0 = 67.11 \: \text{km s}^{-1} \: \text{Mpc}^{-1}$, $\sigma_8 = 0.834$ and $n_{\rm s} = 0.9624$. Each box has size 1 (Gpc/h)$^3$, and is evolved from random Gaussian initial conditions at $z = 127$ to $z = 0$ using the \texttt{GADGET-III} code~\citep{Springel}. We make use of the high-resolution suite, which contains $1024^3$ particles, giving a particle mass of
$8.207\times10^{10} \: h^{-1}M_\odot$. The boxes differ purely in the random phases of the initial conditions, effectively corresponding to simulation of a different random patch of the Universe. We fit each of these boxes rather than just one in order to investigate the sensitivity of our results to the initial conditions---ideally the HMF functional form would not depend on this. Haloes are identified using the Friends-of-Friends (FoF) halo finder~\citep{FOF} with linking length $b=0.2$, and the mass is the sum of all bound particle masses.
Fiducially we use mass bins of width 0.2 dex in the range $\log(M_h/h^{-1}M_\odot)=12.6$--$15.8$ (16 bin centres at $\log(M_h/h^{-1}M_\odot) = 12.7, 12.9, \ldots, 15.7$, although in 38 realisations the highest-mass bin is dropped due to containing zero haloes). The lowest bin edge corresponds to haloes of $\sim50$ particles, sufficient for FoF mass determination~\citep{Jenkins_2001}.
In Appendix~\ref{app:hmf_full} we present results for an extended mass range down to bin centre 12.3 (two further 0.2 dex bins, down to $\sim19$ particles per halo) as a sensitivity analysis.

\section{Method}
\label{sec:method}

\subsection{Exhaustive Symbolic Regression}
\label{sec:ESR}

Symbolic regression (SR) describes the search for the optimal functional forms with which to describe a data set. The space of possible functions is vast and one normally resorts to some form of machine learning algorithm which is stochastic, such as reinforcement learning~\citep{Petersen_2021,Landajuela_2022,Tenachi_2023,Biggio_2021}, Markov Chain Monte Carlo~\citep{Jin_2019}
or genetic programming~\citep{turing,David, haupt,pysr}.
In the latter approach, the most popular, trial functions are generated probabilistically by mutating and cross-breeding functions which have already been identified as giving high likelihood to the data. Genetic programming can survey large swathes of function space but has a tendency to converge on highly complex functions without any guarantee that they are optimal~\citep{ESR,Kronberger}.

Given that the functions that have previously been used to model the LF, SMF and HMF are quite simple
it makes sense to use a systematic, exhaustive approach that considers all possible functions up to a maximum complexity, which we define to be the number of operators, variables and parameters appearing in the function. This can be done using Exhaustive Symbolic Regression (ESR) which has been shown to be remarkably effective at determining the optimal functions at low complexity in a range of different contexts~\citep{ESR,ESR_on_RAR,ESR_priors,Sousa_2023,Martin_2,Martin_1}.

The procedure of ESR is as follows (for full details see~\citealt{ESR}). We first establish a basis set of operators from which we build functions, in our case
\begin{itemize}[nosep]
\item \textbf{Nullary:} $x$, $\theta$ (data variable and free numerical parameter)
\item \textbf{Unary:} $\exp$, $\ln$, inv
\item \textbf{Binary:} $+$, $-$, $\times$, $/$, $\mathrm{pow}$.
\end{itemize}
(pow and log implicitly take the absolute value of their argument to prevent domain errors.) We then use these operators to construct functions of a given complexity.
We start at the lowest complexity that allows nontrivial functions ($3$), and then increase it until we reach the limit of computationally feasibility (10). At each complexity, ESR performs a series of simplifications to identify the unique functions composed of the desired operators, greatly reducing the number of functions that must be fitted to the data.

After the functions have been generated, simplified and de-duplicated they must be evaluated on the data in question. Our approach is
based on the Minimum Description Length (MDL) principle~\citep{RISSANEN1978, MDL_review1, MDL_review2, ESR, ESR_priors}. This takes into account both the complexity of the function and how well it fits the data, collapsing the usual two-dimensional optimisation procedure (over accuracy and simplicity separately) into an unambiguous one-dimensional one. The description length (DL) is the number of nats needed to communicate the data with the aid of the expression, and may be calculated as \citep{ESR}
\begin{equation}
    \label{eq:mdl}
    \begin{split}
        DL\equiv&-\ln({\hat {\cal L}})+k\ln(n)+\sum_j\ln(c_j)-\frac{p}{2}\ln(3) \\
        &+\sum_i^p\left[\frac{1}{2}\ln({\hat I}_{ii})+\ln(|{\hat \theta}_i|)\right],
    \end{split}
\end{equation}
where $D$ is the data being fitted, ${\cal L}$ the likelihood, $\theta_i$ the $i$th free parameter in the function, $p$ the number of free parameters, $k$ the function's complexity,
$n$ the number of unique operators involved, $I$ the observed Fisher information matrix of the parameters and $c_j$ the $j$th constant natural number produced by simplifications. A hat denotes evaluation at the maximum-likelihood point, which is identified using \texttt{BFGS} optimisation~\citep{BFGS_1,BFGS_2} over the parameters of each function. The first term of Eq.~\ref{eq:mdl} describes the accuracy, the second and third terms the structural complexity, and the fourth and fifth terms the parametric complexity. The DL approximates the negative log Bayesian evidence~\citep{Priors} giving it a probabilistic interpretation: the relative probability of function $i$ is proportional to $\exp(-DL_i)$.
In particular it uses the Laplace approximation with uncorrelated parameter values (an assumption we test explicitly in Sec.~\ref{sec:disc}) with a $1/\theta$ parameter prior. This is one level of sophistication beyond the Bayesian Information Criterion, which marginalises parameters out assuming many fewer parameters than data points.

\subsection{Application to the LF, SMF and HMF}
\label{sec:ESR_app}

We denote the LF, SMF and HMF as $\phi(L)$, $\phi(M_\star)$ and $n(M_h)$, respectively, giving the abundance of objects per unit volume and dex of the respective galaxy/halo quantity.
We adopt the common approach of parametrising the HMF through the mass variance $\sigma(M_h)$ associated with haloes of mass $M_h$:
\begin{equation}\label{eq:f_sigma}
n(M_h) \mathrm{d}\log(M_h) = f(\sigma) \: \frac{{\bar \rho}_{\rm m}}{M_h} \: |\mathrm{d}\ln(\sigma)|,
\end{equation}
where
\begin{equation}
\sigma^2(M_h) = \frac{1}{2\pi^2} \int_0^\infty P(k) \: {\tilde W}^2 (kR) \: k^2 \: \mathrm{d}k,
\end{equation}
for linear matter power spectrum $P(k)$, Fourier transform ${\tilde W}$ of the top hat window function of radius $R$ ($\tilde{W}(kR)=3(\sin(kR)-kR\cos(kR))/(kR)^3$) and mean matter density ${\bar \rho}_{\rm m}$. The smoothing scale $R$ is related to the halo mass by $R=(3M_h/(4\pi\bar{\rho}_{\rm m}))^{1/3}$.
This parametrisation is useful because most of the redshift and cosmology dependence of the HMF is already captured by $\sigma$, so $f(\sigma)$ is more universal than $n(M_h)$ \citep{Jenkins_2001, Sheth_Tormen_1999}.

We assume uncorrelated Poisson uncertainties on the counts within each bin of luminosity, stellar mass or halo mass (another assumption we test explicitly in Sec.~\ref{sec:disc}).
The likelihood of observed count $N$ in a given bin is therefore given by
\begin{equation}\label{eq:L_poisson}
\mathcal{L}(N|\lambda) = \frac{\lambda^N \exp(-\lambda)}{N!},
\end{equation}
where $\lambda$ is the expectation from the function evaluated at the bin centre. For luminosity and stellar mass we symbolically regress $\phi(L)$ or $\phi(M_\star)$, i.e. the LF and SMF themselves. After generation, the functions are multiplied by $V_\text{eff}(M)$ (see Sec.~\ref{sec:LF_SMF}) to produce $\lambda$, and the likelihood evaluated using Eq.~\ref{eq:L_poisson}. For halo mass, we instead symbolically regress the $f(\sigma)$ that appears in Eq.~\ref{eq:f_sigma}. We calculate $\sigma$ at the centre of each \texttt{Quijote} mass bin using the \texttt{hmf} module \citep{HMFcalc,hmf}, as well as on a finer grid to evaluate $\mathrm{d}\ln(\sigma)/\mathrm{d}\log(M_h)$ using a cubic spline to calculate $n(M_h)$ through Eq.~\ref{eq:f_sigma}.
To obtain $\lambda$ we then multiply by $V_\text{eff}$, which in this case is just the total box volume $(1 \text{Gpc}/h)^3$, because all objects are included in the simulation. Specifically, the data variable for ESR ($x$) is $L/10^9 L_\odot$, $M_\star/10^9 M_\odot$ or $\sigma$.

We fit ESR directly to two forms of each of the LF and SMF (S\'ersic and cmodel), and ten \texttt{Quijote} realisations of the HMF (those labelled 0, 10, 20, etc.).
To determine how functions perform across the full range of \texttt{Quijote} realisations, we identify the top 200 functions across the 10 explicitly-fitted realisations by combining the description lengths. Specifically, we rank all unique functions by DL within each realisation, average their ranks across realisations, and select the 200 functions with the lowest mean rank. This reduced function set is then fitted to all realisations, whereby we highlight those that frequently achieve ranks within the top five. This shows which functions consistently perform well across all realisations, i.e. are robust to the random variation in initial conditions across the \texttt{Quijote} suite.

We compare our functions to literature standards to assess their quality relative to established fitting functions. The LF and SMF are most commonly described by the \emph{Schechter function}~\citep{Schechter}:
\begin{equation}
\phi(x) = \frac{\theta_0}{\theta_1}\left(\frac{x}{\theta_1}\right)^{\theta_2} e^{-x/\theta_1},
\label{eq:schechter}
\end{equation}
where $x \equiv L/10^9\,L_\odot$ (or $M_\star/10^9\,M_\odot$ for the SMF), $\theta_0/\theta_1 \equiv \phi^*$ is the normalisation, $\theta_1 \equiv L^*/10^9\,L_\odot$ (or $M^*_\star/10^9\,M_\odot$) is the characteristic luminosity (or mass) at which the exponential cutoff sets in, and $\theta_2 \equiv \alpha$ is the faint-end (or low-mass) slope.
Since the Schechter function rises as a power law at small $x$ (for $\alpha < 0$), a double Schechter form is often preferred where there is evidence for a low-mass turnover~\citep{Baldry2012, Wright2017}:
\begin{equation}\label{eq:dbl_schechter}
    \phi(x) = e^{-x/x_*}\left[\phi_1(x/x_*)^{\alpha_1} + \phi_2(x/x_*)^{\alpha_2}\right]/x_*,
\end{equation}
with five free parameters. We fit this form to all four LF/SMF datasets alongside the single Schechter.
A more complex and accurate form \citep{Bernardi} is given by:
\begin{equation}
x\,\phi(x) = \phi_\alpha \beta \left(\frac{x}{x_*}\right)^{\!\alpha} \frac{e^{-(x/x_*)^\beta}}{\Gamma(\alpha/\beta)} + \phi_\gamma \left(\frac{x}{x_\gamma}\right)^{\!\gamma} e^{-x/x_\gamma},
\label{eq:bernardi}
\end{equation}
where $\phi_\alpha$, $x_*$, $\alpha$, $\beta$, $\phi_\gamma$, $x_\gamma$ and $\gamma$ are free parameters (seven in total) and $\Gamma$ is the Gamma function.
This parametrisation enforces positive-definiteness: since $\phi_\alpha, \beta, \Gamma(\alpha/\beta), \phi_\gamma > 0$, both terms are strictly positive and $\phi(x) > 0$ everywhere.
For fitting purposes, we reparametrise Eq.~\ref{eq:bernardi} as:
\begin{equation}\label{eq:bernardi_2}
\phi(x) = \theta_0\, x^{\theta_1} e^{-\theta_2\,x^{\theta_3}} + \theta_4\, x^{\theta_5} e^{-\theta_6\,x}.
\end{equation} This absorbs the $\Gamma$, $x_*$ and $x_\gamma$ factors into the free coefficients: $\theta_0 = \phi_\alpha \beta\, x_*^{-\alpha} / \Gamma(\alpha/\beta)$, $\theta_1 = \alpha - 1$, $\theta_2 = x_*^{-\beta}$, $\theta_3 = \beta$, $\theta_4 = \phi_\gamma\, x_\gamma^{-\gamma}$, $\theta_5 = \gamma - 1$, and $\theta_6 = 1/x_\gamma$. We allow all seven $\theta_i$ to vary independently, spanning a strictly larger function family than Eq.~\ref{eq:bernardi}: the signs of $\theta_0$ and $\theta_4$ are unconstrained, so the reparametrised form can represent functions in which the two components partially cancel, allowing $\phi < 0$. As we will discuss in Sec.~\ref{sec:results_LF_SMF}, the optimiser exploits this freedom in all four datasets, finding fits that are genuinely different from Bernardi's positive-definite form. Henceforth, ``Bernardi function'' refers to Eq.~\ref{eq:bernardi_2} unless otherwise stated; Eq.~\ref{eq:bernardi} is instead referred to as the ``original'' form.

For the HMF, fitting functions commonly used for $f(\sigma)$ include (respectively \citealt{Press-Schechter,Warren_HMF,Tinker_HMF}):
\begin{subequations}\label{eq:f_sigmas}
    \begin{eqnarray}
    f_\text{P.--Sch.}(\sigma) &=& \sqrt{\frac{2}{\pi}} \frac{\delta_c}{\sigma} \exp\left(-\frac{\delta_c^2}{2\sigma^2}\right), \label{eq:f_ps} \\
    f_\text{Warren}(\sigma) &=& \theta_0 (\sigma^{\theta_2} + \theta_1) \exp\left(-\frac{\theta_3}{\sigma^2}\right), \label{eq:f_warren} \\
    f_\text{Tinker}(\sigma) &=& \theta_0\left[\left(\frac{\sigma}{\theta_2}\right)^{-\theta_1}\!\!+1\right]e^{-\theta_3/\sigma^2}, \label{eq:f_tinker}
    \end{eqnarray}
\end{subequations}
where $\delta_c = 1.686$ is the linearly extrapolated density threshold above which structures collapse and the $\theta_i$ are again free parameters.
Note that the Warren and Tinker forms are algebraically equivalent: both reduce to $(\mathcal{C}_1\,\sigma^\alpha + \mathcal{C}_2)\,\exp(-c/\sigma^2)$ with four free parameters (mapping $\mathcal{C}_1 = \theta_0$, $\alpha = \theta_2$, $\mathcal{C}_2 = \theta_0\theta_1$, $c = \theta_3$ for Warren and $\mathcal{C}_1 = \theta_0\theta_2^{\theta_1}$, $\alpha = -\theta_1$, $\mathcal{C}_2 = \theta_0$, $c = \theta_3$ for Tinker). However, the parametrisations differ in an important respect: Tinker's additive ``$+1$'' forces $f(\sigma) > 0$ for all $\sigma$ when $\theta_0 > 0$, whereas Warren's additive $\theta_1$ is a free parameter that can take negative values, permitting solutions where $f$ changes sign. The two forms therefore access different regions of parameter space and can converge to different best-fit solutions.
We fit each of these functions to our datasets and calculate their DL's to locate them within the full function rankings.

In addition to ranking functions by description length, we subject functions to a set of ``physicality checks'' to assess whether they behave sensibly beyond the data range. These are:
\begin{itemize}
	\item \emph{Asymptotic behaviour}---whether $f(x) \to 0$ as $x \to \infty$ (requiring that the abundance of very massive or luminous objects vanishes) and $f(x)$ remains finite and non-negative as $x \to 0^+$.
	\item \emph{Monotonicity}---whether $f(x)$ is monotonically decreasing over the data range, as a consistency check for the single-component, relatively low-complexity models under consideration. We note that monotonicity is not a fundamental physical requirement: real galaxy LFs and SMFs may show non-monotonic structure, though the steepening faint-end slope and high-mass shoulder that motivate the double-component Bernardi form are strictly gradient changes rather than true non-monotonicities. A bump within the data range could indicate a fitting artefact for the low-complexity functions tested here, while non-monotonicities occurring outside the data range may simply reflect unconstrained extrapolation behaviour. The monotonicity check is not relevant for the HMF, for which $f(\sigma)$ peaks at $\sigma \approx 1$.
	\item \emph{Integrability}---whether the integrated mass or luminosity converges to a physically reasonable total density.
    For the SMF, we require $\rho_\star = (10^9 M_\odot / \ln 10) \int \phi(M_\star)\,\mathrm{d}(M_\star/10^9 M_\odot) < \Omega_b \rho_\mathrm{crit} \approx 6.2 \times 10^9 \, M_\odot \, \mathrm{Mpc}^{-3}$~\citep{Planck18_Parameters}, i.e.\ that the total stellar mass density does not exceed the cosmic baryon budget. For the LF, we check that the total luminosity density is consistent with observed values ($\rho_L \sim 10^8 \, L_\odot \, \mathrm{Mpc}^{-3}$; \citealt{Blanton_2003}). For the HMF, the physically relevant integral is the collapsed mass fraction $\int f(\sigma)\,\mathrm{d}\ln\sigma$, which must converge to a value $\leq 1$; we check both convergence and this bound. Integrals are computed over $x \in [10^{-4},\,10^4]$ using $2 \times 10^5$ logarithmically spaced evaluation points, and convergence is verified by extending the integration range by a factor of~10 on each side.
\end{itemize}

A full run of ESR to complexity 10 as used here takes $\sim$700 CPU-hours on Intel Xeon E5-2680 v4 / Intel Xeon Silver 4214 processors.

\section{Results}
\label{sec:results}

\begin{figure*}
\includegraphics[width=\textwidth]{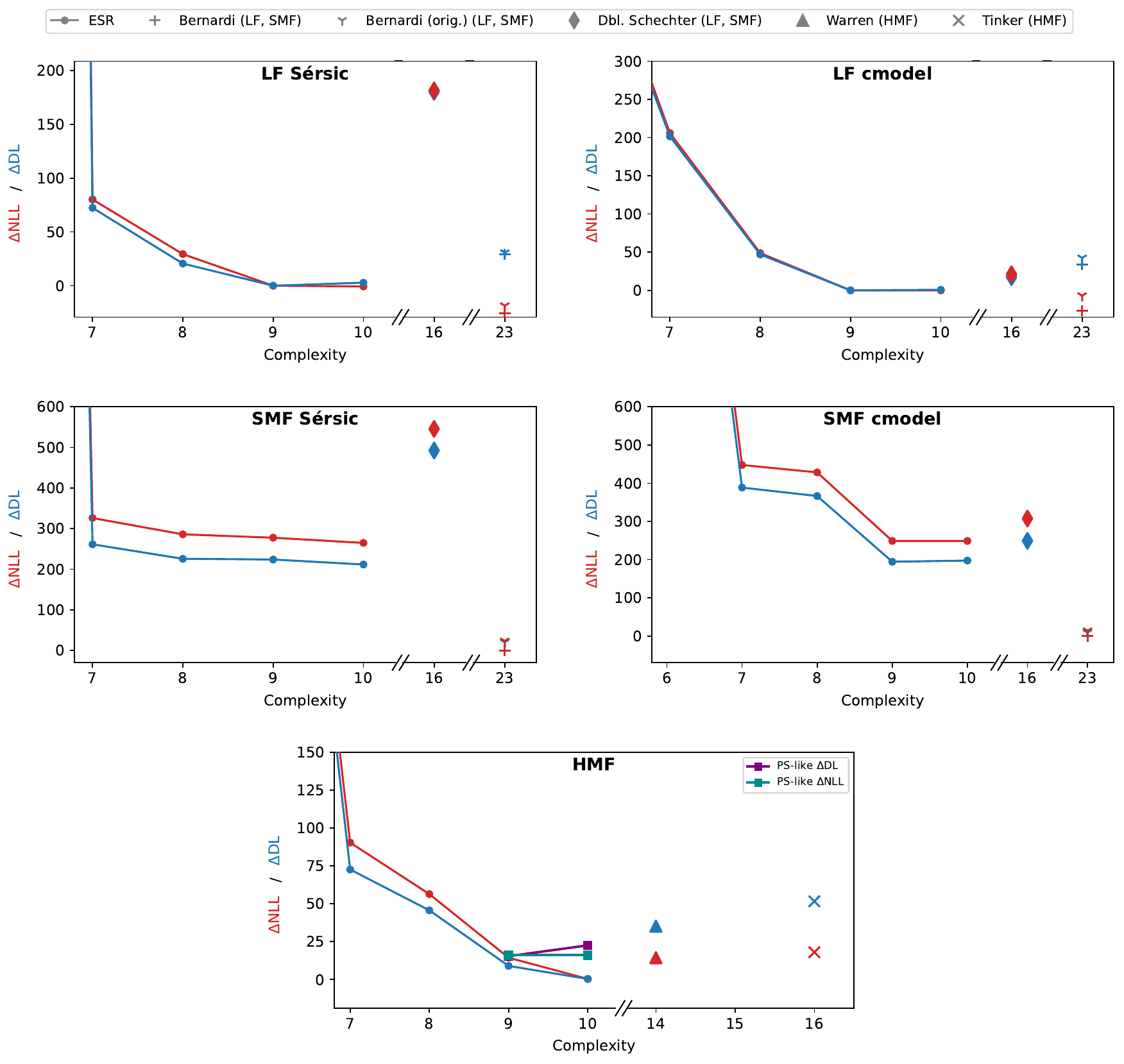}
\caption{
Pareto fronts of negative log-likelihood ($\Delta$NLL, red) and description length ($\Delta$DL, blue) relative to the best function identified, as a function of complexity, for each dataset. Lower is better. Points connected by lines show the best ESR functions at each complexity; isolated markers show the literature fitting functions. We show both the original form of the Bernardi function (Eq.~\ref{eq:bernardi}) and our reparametrisation (Eq.~\ref{eq:bernardi_2}). Complexity gaps are indicated by broken $x$-axes. The best DL functions clearly Pareto-dominate all literature fits besides the Bernardi function for the SMF, indicating a clear gain from using symbolic regression. Note that the Schechter results for the LF and SMF, and Press--Schechter results for the HMF, lie off the top of the plot. The HMF panel additionally shows the Pareto front of Press--Schechter-like (PS-like) functions ($f(\sigma) \propto 1/\sigma$ at large $\sigma$) at the two complexities where we find them.
}
\label{fig:Pareto}
\end{figure*}

\subsection{Galaxy luminosity and stellar mass functions}
\label{sec:results_LF_SMF}

Fig.~\ref{fig:Pareto} shows the Pareto fronts for the LF and SMF S\'ersic and cmodel datasets. The ESR points show the lowest DL (blue) and the negative of the maximum log-likelihood (NLL, red) across all functions at each complexity, while the symbols show the same for the literature functions. The DL and NLL are defined relative to the best (i.e. lowest) values identified, which therefore appear at 0. For the LF and SMF, these decrease sharply between complexities 6 and 8, with diminishing returns at higher complexity, indicating that such complexities are sufficient to reproduce the data well. The Schechter function sits well above the ESR Pareto front (off the top of the plots, hence not shown) despite having comparable complexity, confirming that ESR identifies superior functional forms. For the LF, the Bernardi function is similarly beaten by ESR at lower complexity, while for the SMF no ESR function performs as well as the Bernardi function, as we discuss further below.

\begin{figure*}
\includegraphics[width=\textwidth]{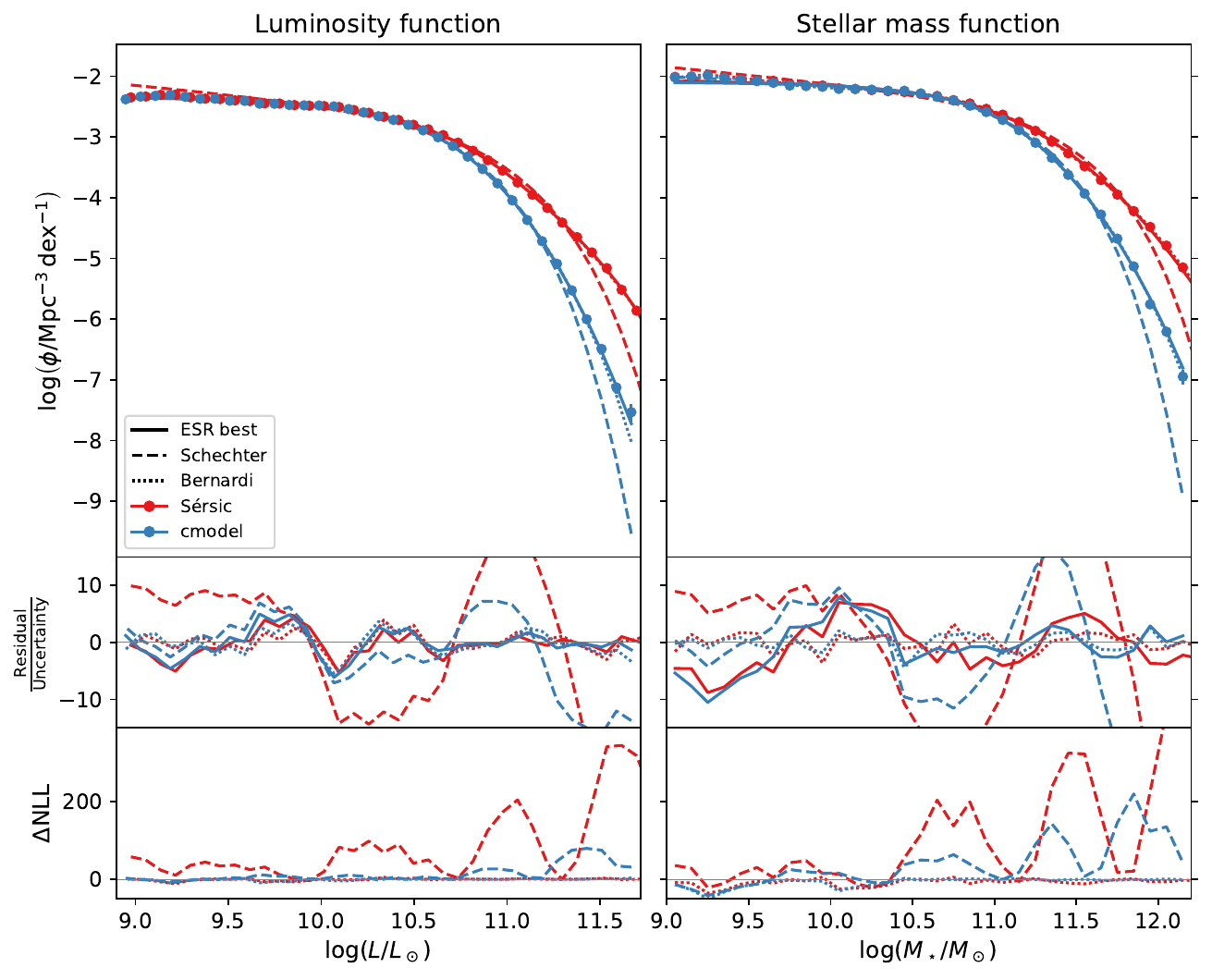}
\caption{Comparison of the best ESR, Schechter and Bernardi fits to the LF (left) and SMF (right) data, for both the S\'ersic (red) and cmodel (blue) photometries.
The upper panels show the data and fits, the middle panels show the uncertainty-normalised residuals, and the lower panels show the per-bin $\Delta$NLL contributions relative to the best ESR function which is therefore a flat line at 0 by construction (not shown). The errorbars on the upper panels
show the asymmetric 68 per cent Poisson confidence interval ($16^\mathrm{th}$--$84^\mathrm{th}$ percentiles) on the count in each bin, converted to $\log\phi$ (these are typically very small). The middle panel uses the symmetric Gaussian approximation to the Poisson uncertainties: $\sigma_{\log{\phi}} = 1/(\ln{10}\sqrt{N})$.
}
\label{fig:LF_SMF}
\end{figure*}

Fig.~\ref{fig:LF_SMF} shows the best functions from ESR for the LF (left) and SMF (right) for both S\'ersic and cmodel photometries. These are compared to the corresponding Schechter and Bernardi fits. We see that the ESR functions perform well across the accessible luminosity range, while the Schechter function underpredicts the abundance at the bright end, and, in the S\'ersic case, overpredicts it at the faint end.

Tables~\ref{tab:LF_functions} and~\ref{tab:SMF_functions} list the top-ranked unique\footnote{While ESR implements an automated symbolic duplicate-removal step it is not foolproof and remaining pairs of functions differing only by reparametrisations ought to be manually removed post-hoc, as we do here.} ESR functions for the LF and SMF respectively, ranked by DL across all complexities up to 10, along with the Schechter and Bernardi fits and the results of physicality checks.
Almost all ESR functions and the Schechter function pass the physicality checks: they vanish at large $x$, remain non-negative, and yield converged integrals corresponding to physical total densities. A few exceptions among the ESR functions are noted. Several functions contain small additive offsets ($\sim\!10^{-5}$) that cause $f(x)$ to become slightly negative at large $x$, producing a divergent integral (note~(a) in the tables). Others are mildly non-monotonic (note~(e)): the rank~1 SMF S\'ersic function increases by $\sim\!7\%$ from $x = 0.01$ to $x \approx 1.4$ due to a cusp at $x = |\theta_1|$, while certain LF S\'ersic functions show similar features. These non-monotonicities occur at or below the low-mass edge of the data range and are therefore extrapolation artefacts rather than data-driven features; they do not cause problems with the integral or asymptotic limits and the affected functions are retained in the ranking. For the SMF, the implied stellar mass densities from all passing functions are $\rho_\star / (\Omega_b \rho_\mathrm{crit}) \approx 0.044$--$0.056$, consistent with the observed cosmic stellar mass fraction~\citep[e.g.][]{Madau_2014}. For the LF, the total luminosity density is $\rho_L \approx 5$--$7 \times 10^7 \, L_\odot \, \mathrm{Mpc}^{-3}$, within observational bounds~\citep{Blanton_2003}. In practice, the data range dominates the total density integral: for the LF, the data range contributes $\sim\!97\%$ of $\rho_L$, and for the SMF $\sim\!99\%$ of $\rho_\star$, so that a good fit to the data ensures a reasonable integral for any function that extrapolates sensibly beyond the data range.

The Bernardi function fails two physicality checks in our fits, but this is a consequence of our reparametrisation. The original form (Eq.~\ref{eq:bernardi}) is a sum of two positive terms and is positive-definite by construction; when refitted under our Poisson likelihood it passes every check, at a modest cost in fit quality. Our reparametrised form (Eq.~\ref{eq:bernardi_2}) releases the sign of $\theta_4$, and the optimiser exploits this across all four datasets to produce fits in which the two components partially cancel, $\phi$ becomes negative at small $x$, and the integral diverges. Refitted shape parameters for the original form agree with \citeauthor{Bernardi}'s published values to within a few per cent across the data range.

\begin{table*}
\centering
\caption{Top ESR functions for the luminosity function, ranked by description length (DL) for complexities up to 10. $x \equiv L / 10^9 \, L_\odot$. $\Delta$DL and $\Delta$NLL are relative to the rank-1 ESR function. The ``Checks'' column summarises the results of the physicality checks described in Sec.~\ref{sec:method} (see notes below the table). Grayed-out rows indicate functions failing at least one check.
}
\label{tab:LF_functions}

\begin{tabular}{>{\rowstyle}c>{\rowstyle}l>{\rowstyle}c>{\rowstyle}c>{\rowstyle}c>{\rowstyle}c>{\rowstyle}c>{\rowstyle}c>{\rowstyle}c>{\rowstyle}c}
\hline
Rank & Function & Comp. & $\theta_0$ & $\theta_1$ & $\theta_2$ & $\theta_3$ & $\Delta$DL & $\Delta$NLL & Checks \\
\hline \hline
\multicolumn{10}{c}{\textbf{LF: S\'ersic}} \\ \hline \noalign{\gmark}
1 & $\theta_0 + |\theta_1|^{|\theta_2|^{|\theta_3|^{x}}}$ & 9 & $-1.27 \times 10^{-5}$ & $1.25 \times 10^{-5}$ & 0.478 & 0.991 & 0 & 0 & (a) \\ \noalign{\nmark}
2 & $|\theta_0|^{|\!\ln|\theta_1 + |\theta_2|^x||^{\theta_3}}$ & 10 & $-2.33 \times 10^{-3}$ & 0.985 & 1.028 & 0.302 & 7.9 & 10.3 & \checkmark \\ \noalign{\gmark}
3 & $|\theta_0 + |\!\ln x|^{x^{\theta_1}}|^{\theta_2}$ & 10 & 18.8 & 0.224 & $-1.84$ & --- & 13.0 & 17.7 & (e) \\ \noalign{\nmark}
4 & $|\!\ln|\theta_0| + |\theta_1|^{x^{\theta_2}}|^{\theta_3}$ & 10 & $3.16 \times 10^{24}$ & 4.28 & 0.314 & $-1.32$ & 13.5 & 15.3 & \checkmark \\ \noalign{\gmark}
5 & $\frac{\theta_0}{-\theta_3 + e^{|\theta_1 - x|^{\theta_2}}}$ & 10 & 0.109 & 1.37 & 0.391 & $-23.7$ & 17.6 & 20.1 & (e) \\ \noalign{\nmark}
6 & $\frac{\theta_0}{\theta_1 + e^{x^{\theta_2}}}$ & 8 & 0.115 & 23.9 & 0.392 & --- & 20.6 & 29.4 & \checkmark \\
7 & $|\theta_0 + \ln|\theta_1 + |\theta_2|^x||^{\theta_3}$ & 10 & 3.00 & 0.339 & 1.033 & $-4.54$ & 21.2 & 18.5 & \checkmark \\ \noalign{\gmark}
8 & $-\theta_3 + |\theta_0|^{\theta_1 - |\theta_2|^{-x}}$ & 10 & $2.74 \times 10^{-4}$ & 1.66 & 1.004 & $1.73 \times 10^{-6}$ & 21.4 & --- & (a) \\
\hline \noalign{\nmark}
10697 & Schechter (Eq.~\ref{eq:schechter}) & 10 & 0.122 & 53.8 & $-0.289$ & --- & 3087 & 3096 & \checkmark \\
--- & Dbl.\ Schechter & 16 & \multicolumn{4}{c}{(5 parameters)} & 222 & 140 & \checkmark \\
\noalign{\gmark}
--- & Bernardi (Eq.~\ref{eq:bernardi_2}) & 23 & \multicolumn{4}{>{\rowstyle}c}{(7 parameters)} & 29 & $-25$ & (b), (c) \\ \noalign{\nmark}
--- & Bernardi (orig.; Eq.~\ref{eq:bernardi}) & 23 & \multicolumn{4}{c}{(7 parameters)} & 31 & $-18$ & \checkmark \\
\hline \hline
\multicolumn{10}{c}{\textbf{LF: cmodel}} \\ \hline
1 & $|\theta_0 + |\theta_1|^{x^{\theta_2}}|^{\theta_3}$ & 9 & 17.0 & $-1.80$ & 0.455 & $-1.85$ & 0 & 0 & \checkmark \\ \noalign{\gmark}
2 & $|\theta_0|^{e^{e^{e^{\theta_1/(\theta_2-x)}}}}$ & 10 & 0.165 & 107 & $-46.0$ & --- & 2.1 & 5.2 & (a) \\ \noalign{\nmark}
3 & $e^{\theta_0 - |\theta_1|^{x/(\theta_2-x)}}$ & 10 & $-4.42$ & 0.0469 & $-98.8$ & --- & 2.7 & 7.5 & \checkmark \\
4 & $|\theta_0|^{e^{|\theta_1 + \theta_2/x|^{\theta_3}}}$ & 10 & $-4.26 \times 10^{-3}$ & 0.654 & 108 & $-1.22$ & 12.5 & 12.5 & \checkmark \\
5 & $\left(-\theta_3 + \left|\frac{\theta_0}{\theta_1 - x}\right|^{\theta_2}\right)^{-1}$ & 10 & $-156$ & $-235$ & $-11.8$ & $-99.0$ & 14.3 & 10.0 & \checkmark \\
6 & $|\theta_0|^{\ln|\theta_3 + (x/|\theta_1|)^{\theta_2}|}$ & 10 & $4.28 \times 10^{-4}$ & 86.2 & 1.21 & 2.02 & 14.4 & 15.5 & \checkmark \\
7 & $|\theta_0|^{\ln|x - |\theta_1 - x|^{\theta_2}|}$ & 10 & $2.46 \times 10^{-4}$ & $-1.93$ & 1.00 & --- & 15.8 & 19.2 & \checkmark \\
8 & $|\theta_0 + |\theta_1|^{e^{x^{\theta_2}}}|^{\theta_3}$ & 10 & 63.0 & 2.06 & 0.179 & $-1.28$ & 19.6 & --- & \checkmark \\
\hline
4596 & Schechter (Eq.~\ref{eq:schechter}) & 10 & 0.124 & 28.3 & $-0.031$ & --- & 488 & 500 & \checkmark \\
--- & Dbl.\ Schechter & 16 & \multicolumn{4}{c}{(5 parameters)} & 59 & $-20$ & \checkmark \\
\noalign{\gmark}
--- & Bernardi (Eq.~\ref{eq:bernardi_2}) & 23 & \multicolumn{4}{>{\rowstyle}c}{(7 parameters)} & 34 & $-27$ & (b), (c) \\ \noalign{\nmark}
--- & Bernardi (orig.; Eq.~\ref{eq:bernardi}) & 23 & \multicolumn{4}{c}{(7 parameters)} & 49 & $-7$ & \checkmark \\
\hline
\end{tabular}

\smallskip
\noindent \emph{Checks:} \checkmark\, = all physicality checks passed. All such functions yield $\rho_L \approx 5$--$7 \times 10^7 \, L_\odot \, \mathrm{Mpc}^{-3}$, consistent with the observed luminosity density.
(a)~Integral diverges.
(b)~Negative at small $x$.
(c)~Negative integral.
(e)~Non-monotonic.
\end{table*}

\begin{table*}
\centering
\caption{As Table~\ref{tab:LF_functions} but for the stellar mass function. $x \equiv M_\star / 10^9 \, M_\odot$.
}
\label{tab:SMF_functions}

\begin{tabular}{>{\rowstyle}c>{\rowstyle}l>{\rowstyle}c>{\rowstyle}c>{\rowstyle}c>{\rowstyle}c>{\rowstyle}c>{\rowstyle}c>{\rowstyle}c>{\rowstyle}c}
\hline
Rank & Function & Comp. & $\theta_0$ & $\theta_1$ & $\theta_2$ & $\theta_3$ & $\Delta$DL & $\Delta$NLL & Checks \\
\hline \hline
\multicolumn{10}{c}{\textbf{SMF: S\'ersic}} \\ \hline \noalign{\gmark}
1 & $|\theta_0 - e^{|\theta_1 + x|^{\theta_2}}|^{\theta_3}$ & 10 & $-36.7$ & $-1.42$ & 0.304 & $-1.32$ & 0 & 0 & (e) \\ \noalign{\nmark}
2 & $|\!\ln|\theta_0| + |\theta_1|^{x^{\theta_2}}|^{\theta_3}$ & 10 & $2.61 \times 10^{21}$ & 3.05 & 0.299 & $-1.22$ & 12 & 13 & \checkmark \\
3 & $|e^{x^{\theta_1}} + \ln|\theta_0||^{\theta_2}$ & 9 & $9.00 \times 10^{15}$ & 0.305 & $-1.31$ & --- & 13 & 21 & \checkmark \\ \noalign{\gmark}
4 & $\frac{\theta_0}{-\theta_3 + e^{|\theta_1 - x|^{\theta_2}}}$ & 10 & 0.344 & 1.77 & 0.329 & $-42.6$ & 14 & 16 & (e) \\
5 & $\theta_0 + \theta_1\,|\theta_2|^{|\theta_3|^x}$ & 9 & $-1.79 \times 10^{-5}$ & $1.79 \times 10^{-5}$ & 436 & 0.998 & 15 & 14 & (a) \\ \noalign{\nmark}
6 & $|\theta_0 - |\theta_1|^{x^{\theta_2}}|^{\theta_3}$ & 9 & $-61.1$ & 3.32 & 0.295 & $-1.16$ & 16 & 17 & \checkmark \\
7 & $|\theta_0 - e^{(2x)^{\theta_1}}|^{\theta_2}$ & 10 & $-64.7$ & 0.294 & $-1.15$ & --- & 17 & 19 & \checkmark \\ \noalign{\gmark}
8 & $\theta_0 + \theta_1 / |\theta_2|^{|\theta_3|^x}$ & 9 & $-1.73 \times 10^{-5}$ & $1.73 \times 10^{-5}$ & $2.21 \times 10^{-3}$ & 0.998 & 20 & --- & (a) \\
\hline \noalign{\nmark}
7265 & Schechter (Eq.~\ref{eq:schechter}) & 10 & 0.647 & 184.5 & $-0.269$ & --- & 4179 & 4187 & \checkmark \\
--- & Dbl.\ Schechter (Eq.~\ref{eq:dbl_schechter})& 16 & \multicolumn{4}{c}{(5 parameters)} & 321 & 240 & \checkmark \\
\noalign{\gmark}
--- & Bernardi (Eq.~\ref{eq:bernardi_2}) & 23 & \multicolumn{4}{>{\rowstyle}c}{(7 parameters)} & $-212$ & $-265$ & (b), (c) \\ \noalign{\nmark}
--- & Bernardi (Eq.~\ref{eq:bernardi}, orig.) & 23 & \multicolumn{4}{c}{(7 parameters)} & $-194$ & $-242$ & \checkmark \\
\hline \hline
\multicolumn{10}{c}{\textbf{SMF: cmodel}} \\ \hline
1 & $|\theta_0 + |\theta_1|^{x^{\theta_2}}|^{\theta_3}$ & 9 & 26.9 & 1.80 & 0.402 & $-1.44$ & 0 & 0 & \checkmark \\
2 & $|\theta_0 + e^{\theta_1 + x^{\theta_2}}|^{\theta_3}$ & 10 & 14.3 & $-1.63$ & 0.321 & $-1.80$ & 9 & 6 & \checkmark \\
3 & $\frac{\theta_0}{\theta_1 + |\theta_2|^{x^{\theta_3}}}$ & 9 & 0.0973 & 11.2 & 1.39 & 0.514 & 10 & 10 & \checkmark \\
4 & $\frac{1}{\theta_0 + |\theta_1|^{e^{x^{\theta_2}}}}$ & 9 & 122 & 2.13 & 0.155 & --- & 17 & 21 & \checkmark \\
5 & $\frac{\theta_0}{\theta_1 + x + e^{x^{\theta_2}}}$ & 10 & 1.27 & 163 & 0.379 & --- & 25 & 29 & \checkmark \\
6 & $\frac{1}{e^{|\theta_1|^{x^{\theta_2}}} + \ln|\theta_0|}$ & 10 & $3.59 \times 10^{51}$ & 2.20 & 0.175 & --- & 26 & 31 & \checkmark \\
7 & $\frac{\theta_0}{\theta_1 + e^{e^{x^{\theta_2}}}}$ & 9 & 3.14 & 406 & 0.143 & --- & 39 & 45 & \checkmark \\
8 & $\frac{1}{\theta_0 + |\theta_1 + \theta_2 x|^{\theta_3}}$ & 10 & 60.2 & 1.85 & $5.22 \times 10^{-3}$ & 6.93 & 40 & --- & \checkmark \\
\hline
5441 & Schechter (Eq.~\ref{eq:schechter}) & 10 & 0.597 & 92.6 & $-0.0932$ & --- & 1208 & 1217 & \checkmark \\
--- & Dbl.\ Schechter (Eq.~\ref{eq:dbl_schechter})& 16 & \multicolumn{4}{c}{(5 parameters)} & 96 & 18 & \checkmark \\
\noalign{\gmark}
--- & Bernardi (Eq.~\ref{eq:bernardi_2}) & 23 & \multicolumn{4}{>{\rowstyle}c}{(7 parameters)} & $-194$ & $-249$ & (b), (c) \\ \noalign{\nmark}
--- & Bernardi (Eq.~\ref{eq:bernardi}, orig.) & 23 & \multicolumn{4}{c}{(7 parameters)} & $-186$ & $-236$ & \checkmark \\
\hline
\end{tabular}

\smallskip
\noindent \emph{Checks:} \checkmark\, = all physicality checks passed. All such functions yield $\rho_\star / (\Omega_b \rho_\mathrm{crit}) \approx 0.044$--$0.056$, consistent with the observed cosmic stellar mass fraction.
(a)~Integral diverges.
(b)~Negative at small $x$.
(c)~Negative integral.
(e)~Non-monotonic.

\end{table*}

\begin{table*}
\centering
\caption{Top ESR functions for the halo mass function, ranked by combined description length across all 100 \texttt{Quijote} realisations. $\Delta$DL and $\Delta$NLL are the combined values summed over all realisations, relative to the rank-1 function. Tinker and Warren are excluded from the ranking because their complexities are beyond the ESR search range. We also include separately the best PS-like functions. Parameter uncertainties show the $16^\text{th}$--$84^\text{th}$ percentile spread in maximum-likelihood parameter values across the 100 realisations. Grayed-out rows indicate functions failing at least one physicality check.}
\label{tab:HMF_functions}

\begin{tabular}{>{\rowstyle}c>{\rowstyle}l>{\rowstyle}c>{\rowstyle}c>{\rowstyle}c>{\rowstyle}c>{\rowstyle}c>{\rowstyle}c>{\rowstyle}c>{\rowstyle}c}
\hline
Rank & Function & Comp. & $\theta_0$ & $\theta_1$ & $\theta_2$ & $\theta_3$ & $\Delta$DL & $\Delta$NLL & Checks \\
\hline \hline
1 & $\frac{\theta_0}{\theta_1 + e^{\sigma^{\theta_2 + \sigma}}}$ & 10 & $0.83^{+0.02}_{-0.02}$ & $0.48^{+0.07}_{-0.06}$ & $-2.22^{+0.01}_{-0.01}$ & --- & 0 & 0 & \checkmark \\
2 & $|\theta_0|^{|\theta_1|^{\exp\left(|\theta_2 - \sigma|^{\theta_3}\right)}}$ & 10 & $0.52^{+0.01}_{-1.03}$ & $1.73^{+0.02}_{-0.02}$ & $1.55^{+0.01}_{-0.00}$ & $2.17^{+0.02}_{-0.02}$ & 204 & 51 & \checkmark \\
3 & $|\theta_0|^{|\theta_1(\theta_2 + \sigma)|^{\ln\sigma}}$ & 10 & $0.26^{+0.00}_{-0.52}$ & $0.40^{+0.01}_{-0.80}$ & $0.13^{+0.02}_{-0.01}$ & --- & 322 & 494 & \checkmark \\
4 & $|\theta_0|^{\sigma^{\theta_1(\theta_2 - \ln\sigma)}}$ & 10 & $0.26^{+0.00}_{-0.52}$ & $-0.89^{+0.01}_{-0.01}$ & $0.87^{+0.01}_{-0.01}$ & --- & 393 & 822 & \checkmark \\
5 & $|\theta_0|^{\theta_1 - \sigma^{-\theta_2 + \sigma}}$ & 9 & $2.43^{+0.04}_{-0.04}$ & $-0.52^{+0.03}_{-0.02}$ & $2.21^{+0.01}_{-0.01}$ & --- & 430 & 1190 & \checkmark \\
6 & $\theta_0\,e^{-(\sigma|\theta_1|)^{\theta_2 + \sigma}}$ & 10 & $0.58^{+0.01}_{-0.01}$ & $1.16^{+0.02}_{-0.02}$ & $-2.44^{+0.03}_{-0.04}$ & --- & 456 & 289 & \checkmark \\
7 & $|\theta_0|^{\sigma^{\ln(\sigma|\theta_1|^{-1})/\theta_2}}$ & 10 & $0.26^{+0.00}_{-0.52}$ & $2.38^{+0.02}_{-4.74}$ & $1.13^{+0.02}_{-0.01}$ & --- & 798 & 822 & \checkmark \\
8 & $|\theta_0|^{|\theta_1|^{(\theta_2 - e^\sigma)/\sigma}}$ & 10 & $0.76^{+0.00}_{-0.00}$ & $0.74^{+0.00}_{-0.00}$ & $-2.54^{+0.03}_{-0.03}$ & --- & 988 & 899 & \checkmark \\
\hline
\multicolumn{10}{c}{\textbf{PS-like ($\sigma f \to \mathrm{const}$ as $\sigma \to \infty$)}} \\ \hline \noalign{\gmark}
9 & $|\theta_0 - \sigma|^{\theta_1 - \sigma^{\theta_2}}$ & 9 & $-0.98^{+0.01}_{-0.01}$ & $-0.96^{+0.01}_{-0.00}$ & $-2.99^{+0.02}_{-0.02}$ & --- & 1038 & 1880 & (c) \\
\noalign{\nmark}
14 & $|\theta_0|^{\theta_1 - |\theta_2|^{\ln\sigma}}/\sigma$ & 10 & $3.34^{+0.04}_{-0.04}$ & $-0.11^{+0.01}_{-0.01}$ & $0.18^{+0.00}_{-0.35}$ & --- & 1247 & 1816 & \checkmark \\
23 & $|\theta_0|^{|\theta_1 - \sigma|^{\theta_2}}/\sigma$ & 9 & $0.33^{+0.01}_{-0.66}$ & $0.14^{+0.01}_{-0.01}$ & $-1.31^{+0.01}_{-0.01}$ & --- & 1537 & 2022 & \checkmark \\
\hline
$>200$ & P.~Sch. (Eq.~\ref{eq:f_ps}) & 10 & $\delta_c = 1.686$ & --- & --- & --- & $1.14 \times 10^7$ & $1.14 \times 10^7$ & \checkmark \\
\noalign{\gmark}
--- & Warren (Eq.~\ref{eq:f_warren}) & 14 & $7.64^{+0.91}_{-0.99}$ & $-0.92^{+0.01}_{-0.01}$ & $-0.04^{+0.00}_{-0.01}$ & $0.81^{+0.01}_{-0.01}$ & 2225 & 1273 & (a) \\
--- & Tinker (Eq.~\ref{eq:f_tinker}) & 16 & $1.1^{+0.4}_{-0.3} \times 10^{-3}$ & $0.75^{+0.02}_{-0.02}$ & $4440^{+2636}_{-1243}$ & $0.90^{+0.01}_{-0.01}$ & 3072 & 2110 & (b) \\
\hline \noalign{\nmark}
\end{tabular}

\smallskip
\noindent \emph{Checks:} \checkmark\, = all physicality checks passed: $f(\sigma) \to 0$ as $\sigma \to \infty$, $f(\sigma)$ remains finite as $\sigma \to 0^+$, and $\int f(\sigma)\,\mathrm{d}\ln\sigma$ converges to a value $\leq 1$ (the collapsed mass fraction).
(a)~$f(\sigma) < 0$ for $\sigma \gtrsim 7$;
$\int f\,\mathrm{d}\ln\sigma$ is negative.
(b)~$f \to \theta_0 > 0$ as $\sigma \to \infty$, so $f$ does not strictly vanish; however $\int f\,\mathrm{d}\ln\sigma$ converges ($\approx 0.79$).
(c)~$f(\sigma) \to \infty$ as $\sigma \to 0^+$; $\int f\,\mathrm{d}\ln\sigma$ diverges.

\end{table*}

The Bernardi function beats the best ESR function in DL for the SMF ($\Delta\mathrm{DL} \sim -200$) but loses for the LF ($\Delta\mathrm{DL} \sim +30$). With similar complexity penalties in both cases, the difference is driven by the NLL. For the SMF, Bernardi fits systematically better---especially at the low-mass end and around the characteristic mass---suggesting genuine two-component structure (a steepening faint-end slope plus a high-mass shoulder) beyond the reach of any single-component function at complexity $\leq 10$. For the LF the per-bin improvements are smaller and partially cancel, so the complexity penalty favours the lower-complexity ESR functions.

\begin{figure*}
	\includegraphics[width=\textwidth]{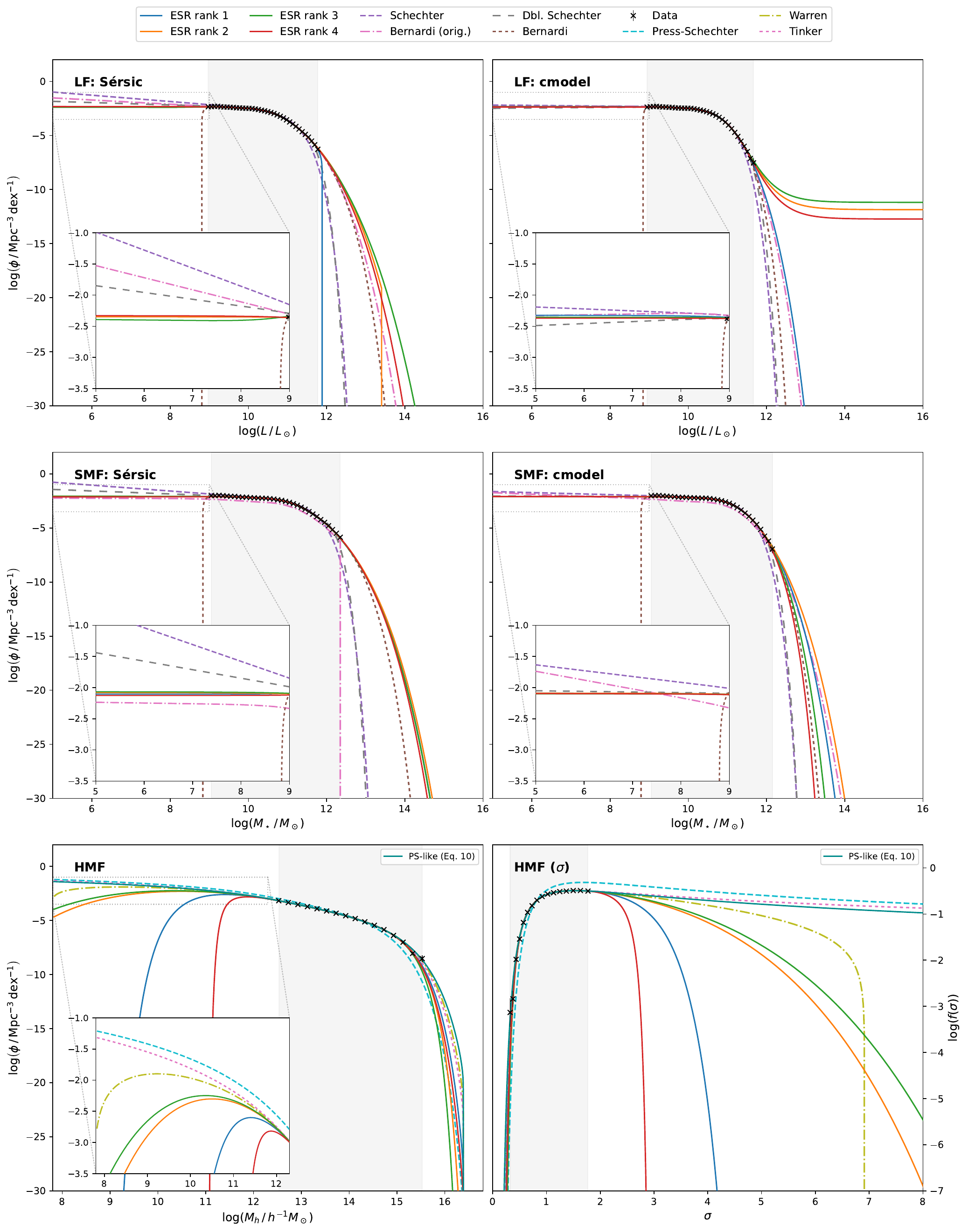}
	\caption{Extrapolation behaviour of the top four ESR functions (solid coloured lines) and literature fits (dashed/dotted lines) for the LF (top), SMF (middle) and HMF (bottom). For the HMF, we show the top four ESR functions from the combined ranking across all 100 \texttt{Quijote} realisations
(the data points and fitted parameter values are for realisation 50, but look similar for any realisation).
    }
	\label{fig:extrap}
\end{figure*}

To visualise the behaviour of the functions under extrapolation, Fig.~\ref{fig:extrap}
plots the top four ESR functions alongside the literature fits for each data set evaluated over a range in luminosity or mass that extends far beyond the observed data (shaded region).
(The Bernardi function goes negative at low luminosities or masses, dropping below $\phi = 0$ and hence disappearing from the logarithmic plot.)
For the LF and SMF, the Schechter function extrapolates smoothly to zero at high luminosities and masses, as expected from its exponential cutoff. However, it rises as a power law at the faint/low-mass end, which, although not leading to a divergent integral since $\alpha > -2$, is discrepant with the notion that galaxy formation is inoperative below a certain mass threshold \citep{Benson2003}.
The double Schechter fit (Eq.~\ref{eq:dbl_schechter}) always beats the single Schechter in NLL (by $500$--$4000$), but always loses to ESR rank~1 in DL (by $59$ to $321$), because the ESR 3--4 parameter forms are more efficient. The second Schechter component manifests as a small-amplitude, steep-slope contribution that activates only at the faint end.

The top-ranked ESR functions extrapolate as a constant as $x \to 0$, which is likely a behaviour encouraged by simplicity rather than the data itself.
For the LF S\'ersic, the rank~1 function has a divergent integral due to a small additive offset, so the rank-2, rank-4 or rank-6 functions, all of which pass all checks, should be considered the best physical models. For the LF cmodel, the rank-1 function passes all checks and extrapolates smoothly to zero at high luminosity. This function shares the same $|\theta_0 + |\theta_1|^{x^{\theta_2}}|^{\theta_3}$ structure as the best SMF cmodel function, suggesting this is a robust functional form for galaxy number counts.

\subsection{Halo mass functions}
\label{sec:results_HMF}

\begin{figure}
	\centering
	\includegraphics[width=\columnwidth]{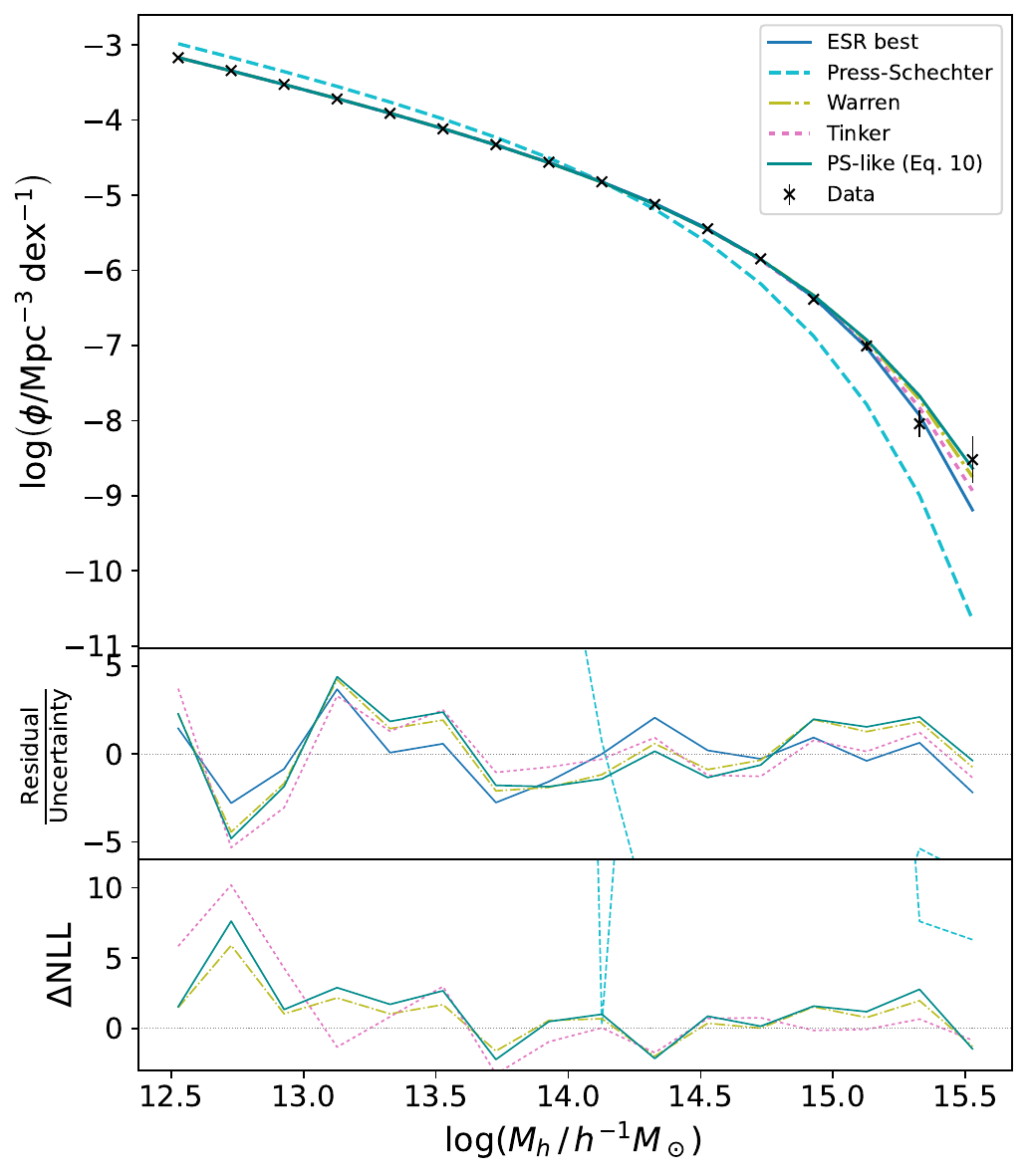}
	\caption{
As Fig.~\ref{fig:LF_SMF} but for the HMF, comparing the top ESR function overall, and best PS-like function, to literature fits. Upper panel: $f(\sigma)$ with data points. Middle panel: uncertainty-normalised residuals. Lower panel: per-bin $\Delta$NLL relative to the rank-1 ESR function. Results are shown for \texttt{Quijote} realisation 50, but are similar across all realisations.
}
	\label{fig:HMF_functions}
\end{figure}

Fig.~\ref{fig:HMF_functions}
shows the analogue of Fig.~\ref{fig:LF_SMF} for the HMF (in this case for \texttt{Quijote} realisation 50).
Here we have only a single mass definition and compare the best ESR functions to the three literature fits given in Eq.~\ref{eq:f_sigmas}. We see a better fit to the data from the ESR functions than the literature ones, with Press--Schechter notably providing a very poor fit as expected.

Table~\ref{tab:HMF_functions} ranks the best ESR functions by their combined description length across all 100 \texttt{Quijote} realisations. Each realisation is fitted independently with its own maximum-likelihood parameters; the combined DL is then
\begin{equation}\label{eq:DL_combined}
	\begin{split}
		\mathrm{DL}_\mathrm{comb} = &\sum_{i=1}^{100}\left[-\ln(\hat{\cal L}_i) - \frac{p}{2}\ln(3) + \sum_j^p\left(\frac{1}{2}\ln(\hat{I}_{jj,i}) + \ln(|\hat{\theta}_{j,i}|)\right)\right] \\
		&+ k\ln(n) + \sum_m\ln(c_m),
	\end{split}
\end{equation}
where the $i$ sum runs over the independent realisations, each with its own likelihood $\hat{\cal L}_i$ and maximum-likelihood parameters $\hat{\theta}_{j,i}$, while the structural complexity terms $k\ln(n) + \sum_m\ln(c_m)$ are counted only once since the functional form is shared.
This explains why the $\Delta$DL values are $\mathcal{O}(100)\times$ larger here than in Tables~\ref{tab:LF_functions} and~\ref{tab:SMF_functions}. For parameter values we quote the mean maximum-likelihood value over all 100 realisations and the $16^\text{th}$--$84^\text{th}$ percentile as the errorbar; the similarity of all realisations makes these uncertainties small.
The most significant contributions to the $\Delta$DL values of Warren and Tinker relative to the best ESR function come from $\Delta$NLL contributions, indicating that ESR improvement is driven by fit quality rather than simplicity.

For the best ESR function, the inter-realisation spread (16th--84th percentile half-width) exceeds the single-realisation Fisher uncertainty by a factor of $\sim\!1.4$--$2.1$ across parameters, indicating that cosmic variance marginally dominates over statistical noise and that the quoted uncertainties are meaningful estimates of the total error budget. For Warren and Tinker the inter-realisation spread exceeds the Fisher uncertainty by factors of $\sim\!30$--$300$ for some parameters, reflecting parameter degeneracies.
Tinker exhibits such a degeneracy between $\theta_0$ and $\theta_2$: in the regime where data constrain the fit ($\sigma \sim 0.3$--$2$), the power-law term $(\sigma/\theta_2)^{-\theta_1}$ dominates the bracket, giving $f \approx \theta_0\,\theta_2^{\theta_1}\,\sigma^{-\theta_1}\,\exp(-\theta_3/\sigma^2)$. The product $\theta_0\,\theta_2^{\theta_1}$ is well constrained but $\theta_0$ and $\theta_2$ individually are not, explaining the broad distributions of both parameters. A similar situation occurs for Warren's $\theta_0$ and $\theta_2$: in the data range where $\sigma^{\theta_2} \gg \theta_1$, the function reduces to $f \approx \theta_0\,\sigma^{\theta_2}\,\exp(-\theta_3/\sigma^2)$, so only the product $\theta_0\,\sigma^{\theta_2}$ is well constrained.
This causes a large spread in best-fit $\theta_0$ values.

Press--Schechter and Tinker both pass the integrability check: although $f$ vanishes only as a power law for Press--Schechter (decaying as $\delta_c/\sigma$) and tends to a small positive constant ($\theta_0 \approx 9 \times 10^{-4}$) for Tinker, the physically relevant integral $\int f\,\mathrm{d}\ln\sigma$ converges in both cases, yielding collapsed mass fractions of 1 and $\approx 0.79$ respectively. Warren, however, fails: $f(\sigma) < 0$ for $\sigma \gtrsim 7$ due to its additive constant $\theta_1 < 0$, causing $\int f\,\mathrm{d}\ln\sigma$ to diverge to $-\infty$.
As noted above, the Warren and Tinker forms are algebraically equivalent but their different parametrisations lead the optimiser to different minima. Warren's free additive constant $\theta_1$ allows solutions with $f < 0$ at large $\sigma$, which fit the data $\sim\!4$ nats better in NLL than any non-negative solution but fail our physicality checks. Tinker's ``$+1$'' structure prevents the optimiser from reaching this regime, confining it to the non-negative subspace where $f \to \theta_0 > 0$.

The large-$\sigma$ behaviour of the literature fits is well understood. Both Warren ($f \to \theta_0\theta_1$) and Tinker ($f \to \theta_0$) tend to a non-zero constant as $\sigma \to \infty$, so $\int f\,\mathrm{d}\ln\sigma$ diverges logarithmically in both cases. On \texttt{Quijote} Warren's best-fit $\theta_1$ is negative, giving the more severe pathology $f < 0$ for $\sigma \gtrsim 7$. Both original references explicitly caution against extrapolating their fits outside the calibrated mass range.

We note that $f(\sigma) \to \mathrm{const}$ at large $\sigma$ does not by itself imply unphysical behaviour in the number density $n(M_h)$: since $n(M_h) \propto f(\sigma)/M_h \cdot |\mathrm{d}\ln\sigma/\mathrm{d}\ln M_h|$ and both $1/M_h \to \infty$ and $|\mathrm{d}\ln\sigma/\mathrm{d}\ln M_h|$ remain finite as $M_h \to 0$, the number density diverges but the \emph{mass} density $\int M_h\,n(M_h)\,\mathrm{d}M_h$ can still converge. This is directly analogous to the LF and SMF, where $\phi \to \mathrm{const}$ at small $L$ or $M_\star$ is not considered pathological. The issue is specific to the \emph{integral} $\int f\,\mathrm{d}\ln\sigma$: if $f \to \mathrm{const} > 0$ as $\sigma \to \infty$, this integral diverges logarithmically, violating the physical requirement that the collapsed mass fraction be $\leq 1$. This is a stronger criterion than normally applied---the literature functions were not designed to satisfy it---but it provides a useful discriminant among extrapolation behaviours.
The extrapolation behaviour of the literature and best ESR functions (separately as a function of $M_h$ and $\sigma$) is shown in Fig.~\ref{fig:extrap}, with inset panel zooming in on the low end of the mass range.

The contrast with the ESR results is instructive. All eight top-ranked ESR functions shown pass the physicality checks, vanishing at large $\sigma$, remaining finite at small $\sigma$, and yielding convergent collapsed mass fractions $\sim 0.36$--$0.47$.
Around 0.28 of this comes from the data range itself, with the remainder coming almost entirely from the low-mass extrapolation.
The success of ESR and relative uniformity of its favoured functional forms suggests that the data themselves, through the description length criterion, indicate certain asymptotic structure beyond the data range. In particular, they settle on a super-exponential cutoff $\sim |\theta_0|^{g(\sigma)}$, implying $f \to 0$ as $\sigma \to \infty$. However, whether $f(\sigma) \to 0$ is physically correct at large $\sigma$ is not settled by our data: zoom simulations extending the HMF to much lower masses find that $f(\sigma)$ continues to rise, approximately consistent with the Press--Schechter $\sim\delta_c/\sigma$ scaling (discussed further in Sec.~\ref{sec:alt_data}). The ESR cutoff should therefore be understood as a property of the optimal fit within the \texttt{Quijote} mass range coupled to the DL metric, rather than as a robust prediction for the HMF behaviour towards arbitrarily low masses.

ESR does, of course, discover functions with PS-like large-$\sigma$ scaling, but they are slightly disfavoured by the data according to DL. Of the top 200 ESR functions evaluated across all 100 realisations, 26 exhibit $f(\sigma) \sim 1/\sigma$ behaviour at large $\sigma$, with the top three listed in Table~\ref{tab:HMF_functions}. After some algebraic manipulation and variable redefinition, the highest-ranked PS-like function that passes all checks, at rank 14 overall, is
\begin{equation}
f(\sigma) = \frac{\alpha \: e^{-\beta \sigma^{\gamma}}}{\sigma}
\label{eq:hmf_ps_like}
\end{equation}
for constants $\alpha, \beta, \gamma$. Transforming the best-fit parameters from Table~\ref{tab:HMF_functions} ($\alpha = |\theta_0|^{\theta_1}$, $\beta = \ln|\theta_0|$, $\gamma = \ln|\theta_2|$) gives $\alpha \approx 0.876$, $\beta \approx 1.206$ and $\gamma \approx -1.715$.
The PS-like Pareto front is close to the unconstrained ESR front in the HMF panel of Fig.~\ref{fig:Pareto}, indicating that PS-like functions---which one may wish to restrict to on theoretical grounds---provide nearly as good a fit as the unconstrained best within our framework.

\begin{figure*}
\centering
\includegraphics[width=1.125\linewidth]{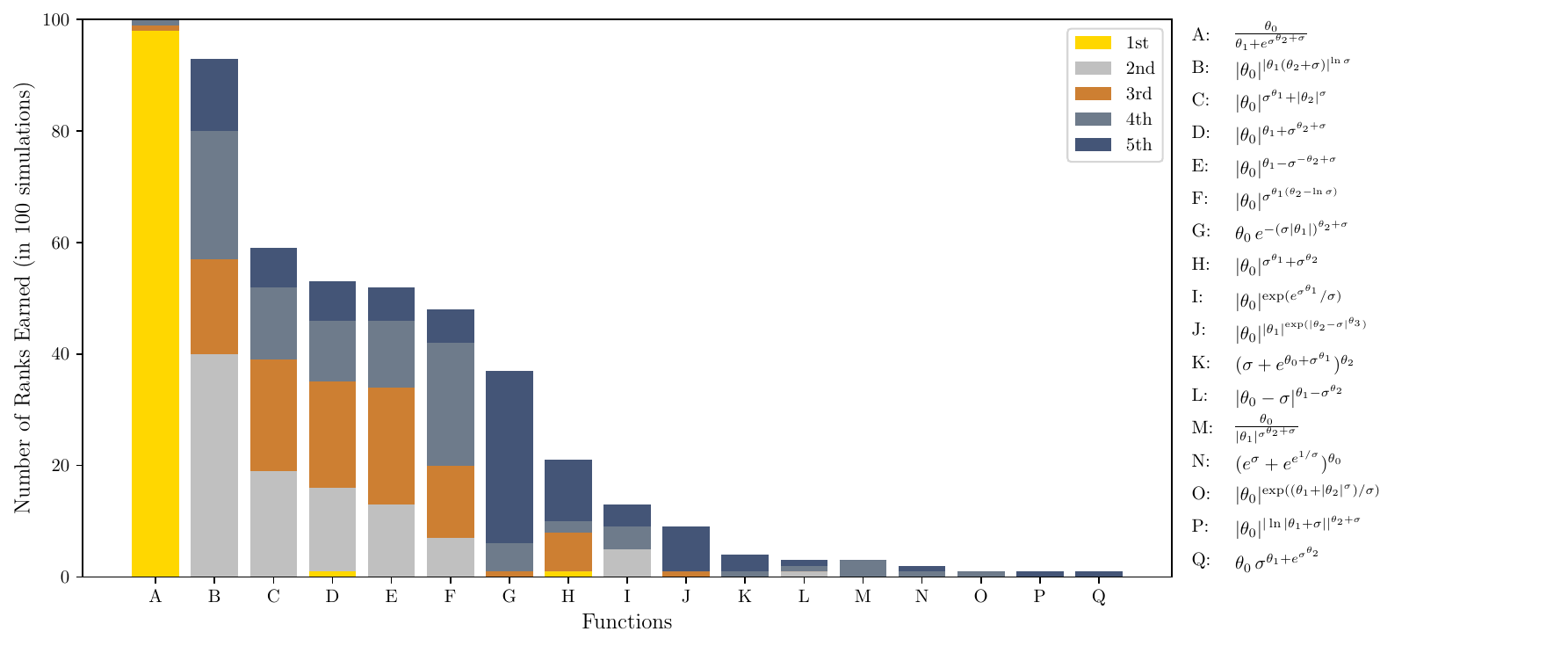}
\caption{Top-five ranks of all functions appearing at least once in the top five of at least one of the 100 \texttt{Quijote} realisations.}
\label{fig:HMF_comp}
\end{figure*}

Finally, as a complement to analysing the top functions on all 100 realisations combined according to Eq.~\ref{eq:DL_combined}, we investigate the number of times individual functions perform well on the realisations separately. This studies the consistency between realisations which the analyses above average over.
Fig.~\ref{fig:HMF_comp} shows the number of times each function appears at each rank in the top five (i.e.\ within the five lowest DL) across the 100 realisations, for each function for which that happens at least once. The functions are ordered by total number of ranks 1--5 earned. We see that the top few functions are consistently good across the realisations; the top-ranked function appears in the top five in all 100 realisations and is ranked first in 98, showing extremely strong dominance across the simulation suite. This gives us confidence (alongside the small variation in best-fit parameter values in Table~\ref{tab:HMF_functions}) that our results are not sensitive to the random phases of the initial conditions across the \texttt{Quijote} suite, and hence that the best ESR functions describe the HMF robustly.

This top-5 frequency ranking is broadly concordant with the combined DL ranking of Table~\ref{tab:HMF_functions}: the same function leads both rankings, and 7 of the top 10 functions appear in both top-10 lists. The two metrics are not identical, however: the combined DL (Eq.~\ref{eq:DL_combined}) sums the NLL over all 100 realisations, rewarding functions that fit \emph{consistently} well, whereas the bar chart counts only whether a function appears in the per-realisation top five, regardless of its performance on the remaining realisations. A function that is occasionally excellent but usually mediocre can therefore rank highly on the bar chart while scoring poorly on the combined DL. For example, function N in Fig.~\ref{fig:HMF_comp} appears in the bar chart (top-5 in 2 of 100 realisations) but sits at combined DL rank~85 because its summed NLL across all realisations is uncompetitive.

\subsection{Overall best functions}
\label{sec:best}

We conclude this section by highlighting the overall best ESR functions for each mass or luminosity function, considering both statistical ranking through DL and physicality criteria.

For the LF, the best ESR functions differ between photometries so no single form can yet be recommended across both. For the S\'ersic photometry, the highest-ranked function that passes all physicality checks is the rank-2 function, which with some algebraic manipulation can be written as
\begin{equation}\label{eq:1}
\phi(x) = e^{\alpha\,\left(\ln\!\left(\beta + \gamma^{\,x}\right)\right)^{\!\delta}},
\end{equation}
The function has complexity~10, four parameters, and $\Delta\mathrm{DL} = 7.9$ relative to rank~1 (Table~\ref{tab:LF_functions}).
As $x \to \infty$, the $\gamma^{\,x}$ term dominates the argument of the logarithm, giving $\phi \sim e^{\alpha\,(x \ln\gamma)^{\delta}} \to 0$ super-exponentially since $\alpha < 0$ and $\delta > 0$. As $x \to 0^+$, $\gamma^{\,x} \to 1$ and $\phi \to e^{\alpha\,(\ln(\beta + 1))^{\delta}}$, a finite constant. With $\beta \approx 0.985$ and $\gamma \approx 1.028$, the function transitions smoothly between these regimes near $x \sim 50$ ($L \approx 5 \times 10^{10}\,L_\odot$).

A simpler alternative is the rank-6 function
\begin{equation}
	\phi(x) = \frac{\theta_0}{\theta_1 + e^{x^{\theta_2}}},
\end{equation}
with complexity~8, only three parameters, and $\Delta\mathrm{DL} = 20.6$ relative to the ESR best.
This logistic-type form provides an exponential cutoff at high luminosity while saturating to a finite constant $\theta_0/\theta_1$ at low luminosity. As $x \to \infty$, $\phi \sim \theta_0 \, e^{-x^{\theta_2}} \to 0$ super-exponentially, ensuring rapid convergence of the luminosity density integral. As $x \to 0^+$, $\phi \to \theta_0 / (\theta_1 + 1)$, a finite constant. Both functions share the same qualitative behaviour---a finite faint-end plateau and a super-exponential bright-end cutoff---but the rank-2 function achieves a better fit to the data
at the cost of one additional parameter and two additional operators.
The Schechter function, by contrast, rises as a power law $\phi \propto x^\alpha$ at low $x$; although its integral also converges (for $\alpha > -2$), the physical plausibility of this continued rise depends on whether galaxy formation is suppressed at very low luminosities. Deeper surveys such as GAMA \citep{Baldry2012, Wright2017} do find a steepening faint-end slope requiring double Schechter forms, which would favour functions with richer low-$x$ structure than the single-plateau ESR form. We therefore caution that the ESR faint-end extrapolation, while well-behaved, is not constrained by data below $L \sim 10^{8.5} \, L_\odot$ and is likely a result of the preference of the DL metric for simplicity.

For the cmodel photometry, the best ESR function can be written as
\begin{equation}\label{eq:good}
\phi(x) = \left(\alpha + e^{\beta x^{\gamma}}\right)^{\!\delta},
\end{equation}
This has complexity~9 and four parameters (Table~\ref{tab:LF_functions}), and is the same power-of-sum form that dominates the SMF results (see below), here with $\delta \approx -1.85$. It passes all physicality checks: as $x \to \infty$, the $e^{\beta x^{\gamma}}$ term dominates and $\phi \sim e^{\delta\beta\, x^{\gamma}} \to 0$ (since $\delta\beta < 0$); as $x \to 0^+$, $\phi \to (\alpha + 1)^{\delta}$, a finite constant. Despite the structural differences between the best S\'ersic and cmodel functions, a common feature emerges: in both photometries, the high-luminosity cutoff is a stretched exponential $\phi \sim e^{-c\,x^p}$ with $c > 0$ and $p \approx 0.3$--$0.5$ (corresponding to the exponent $\delta$ in Eq.~\ref{eq:1} and $\gamma$ in Eq.~\ref{eq:good}), i.e.\ steeper than the simple exponential of the Schechter function within the data range. This suggests that the bright-end cutoff is robustly steeper than exponential, regardless of the photometric model.

For the SMF, the same functional form Eq.~\ref{eq:good}
ranks first for the cmodel photometry and appears in a closely related form (with a minus sign, rank~6) for S\'ersic (Table~\ref{tab:SMF_functions}).
With only four parameters and complexity 9, it is simpler and statistically superior to the Schechter function ($\Delta\mathrm{DL} > 1000$). However the Bernardi function achieves the lowest DL overall for both SMF photometries ($\Delta\mathrm{DL} \approx 194$--$212$ below the best ESR function), indicating that the DL minimum for the SMF lies beyond the ESR range. We therefore cannot provide the statistically optimal fit to the SMF, but can provide principled simpler alternatives to the Bernardi function.
The same caveats about the low-$x$ extrapolation apply as for the LF: the data range contributes $\sim\!99\%$ of $\rho_\star$, so the plateau is effectively unconstrained by data and may not reflect the true behaviour at $M_\star \ll 10^{8.5} \, M_\odot$.

For the HMF, the best function across all 100 realisations is
\begin{equation}\label{eq:2}
f(\sigma) = \frac{\theta_0}{\theta_1 + e^{\sigma^{\theta_2 + \sigma}}},
\end{equation}
a logistic-type form with complexity 10 and only three free parameters ($\theta_0 \approx 0.83$, $\theta_1 \approx 0.48$, $\theta_2 \approx -2.22$). As $\sigma \to \infty$ the exponential in the denominator diverges super-exponentially, giving $f \to 0$; as $\sigma \to 0^+$, $f \to \theta_0/(\theta_1 + e)$, a finite positive constant.
Several of the top ESR functions have a nested exponential motif $\sim|\theta_0|^{g(\sigma)}$, appearing at ranks 2, 3, 4, 7 and 8 in Table~\ref{tab:HMF_functions}.
The top ESR functions fit the data better than
Warren and Tinker with a substantially lower DL overall, driven primarily by a better NLL
and secondarily by a smaller contribution from the lower complexity.
However, the super-exponential cutoff at \emph{large} $\sigma$ (low halo masses) seen in several functions is driven by the functional form rather than by the data, and is unlikely to be realistic as we discuss in Sec.~\ref{sec:alt_data}. The best PS-like function with $1/\sigma$ behaviour at large $\sigma$ is shown in Eq.~\ref{eq:hmf_ps_like}.

\section{Discussion}
\label{sec:disc}

\subsection{Implications of results}
\label{sec:implications}

The Schechter function has been the standard parametrisation of the LF and SMF for nearly five decades, valued for its simplicity and the ease with which its parameters can be compared across surveys, redshifts, and environments. Our results show that ESR can identify functions that fit the data significantly better---by $\Delta\mathrm{DL} > 500$ for the LF and $> 1000$ for the SMF---while using comparable or fewer parameters. However, replacing a community standard requires more than statistical superiority on a single dataset.
We suggest the ESR functions are most useful in contexts where: the fit quality matters (e.g.\ forward-modelling galaxy populations in cosmological analyses where per-cent-level accuracy in the mass function is required; e.g. \citealt{Behroozi2013, Moster2013}), and the bright/massive end is important (where the Schechter function systematically underfits). At least for the SMF, more complex functional forms such as Eq.~\ref{eq:bernardi} appear warranted.

For the HMF, the Press--Schechter formalism and its descendants (e.g. Tinker and Warren) have served as the standard parametrisations of the multiplicity function $f(\sigma)$ for over three decades. Our ESR analysis, applied across 100 \texttt{Quijote} realisations, identifies functions that outperform all three literature forms in combined description length---by $\Delta\mathrm{DL} \approx 2200$--$3100$ relative to Warren and Tinker (Table~\ref{tab:HMF_functions}), and by $\sim\!1.1 \times 10^7$ relative to Press--Schechter, using 3--4 free parameters. The top ESR functions exhibit super-exponential cutoffs at large $\sigma$---whether through a denominator that grows super-exponentially (rank~1: $\theta_0/[\theta_1 + e^{\sigma^{\theta_2+\sigma}}]$) or through the $|\theta_0|^{g(\sigma)}$ motif (ranks~2--4)---representing a qualitatively different functional structure from the power-law $\times$ exponential forms prevalent in the literature.
As with the LF and SMF, we suggest the ESR functions are most useful where fit quality is important---for example, in precision cosmological analyses that require accurate halo abundances for cluster counts \citep[e.g.][]{Rozo2010, Allen2011} or weak-lensing mass calibration \citep[e.g.][]{vonderLinden2014}.

To quantify the practical impact, we estimate the effect of the function choice on key derived quantities. For the SMF, the total stellar mass density $\rho_\star$ computed from the ESR best function is $0.3\%$ lower than the Schechter value for S\'ersic photometry over the data range and identical for cmodel photometry (respectively $2.5\%$ and $0.5\%$ lower when extrapolated to $M_\star = 10^7$--$10^{12}\,M_\odot$). Similarly, for the LF, $\rho_L$ is $1.4\%$ and $0.3\%$ lower for the ESR function over the data range, extending to $4.6\%$ and $0.6\%$ respectively when extrapolated over $L = 10^{7}$--$10^{12}\,L_\odot$. For galaxy counts in an SDSS-like survey volume ($\sim\!10^8\,\mathrm{Mpc}^3$) above $L > 10^{10}\,L_\odot$, the ESR function predicts $6.2\%$ more galaxies than Schechter for the S\'ersic photometry and $1.5\%$ for cmodel. For the HMF, the ESR best function predicts $n(>M_h)$ values that are $0.6\%$ lower than both Warren and Tinker at $M_h \sim 10^{13}\,h^{-1}M_\odot$, and up to $1.3\%$ higher at $M_h \sim 10^{14}\,h^{-1}M_\odot$ ($11.7\%$ lower than Warren and $1.2\%$ higher than Tinker at $M_h > 10^{15}\,h^{-1}M_\odot$). Press--Schechter, by contrast, overpredicts $n(>10^{13}\,h^{-1}M_\odot)$ by $\sim\!29\%$ and underpredicts $n(>10^{15}\,h^{-1}M_\odot)$ by a factor of $\sim\!3.5$ relative to the ESR best function. Thus, within the data range, the function choice has a modest ($\lesssim\!2\%$) impact on integrated quantities, but becomes significant ($5$--$10\%$) under extrapolation and at the extreme high-mass end of the HMF. This shows that the functional form of these functions has a non-negligible impact on astrophysically and cosmologically relevant quantities, making our improved fits important.

A striking result is that ESR independently discovers super-exponential cutoffs in all three contexts---the LF, SMF, and HMF---despite the very different physical regimes and data characteristics involved. The evidence for this cutoff rests on different foundations in each case.
For the LF and SMF, the bright/massive end is well constrained by the data: the SDSS bins at $L \gtrsim 10^{10.5}\,L_\odot$ and $M_\star \gtrsim 10^{10.5}\,M_\odot$ contain tens to thousands of galaxies per bin, so the super-exponential cutoff identified by ESR in this regime is a robust feature of the data rather than an artefact of the DL metric. The same super-exponential structure is in fact preferred across all four LF/SMF datasets (S\'ersic and cmodel photometry).
For the HMF the situation is somewhat different.
The high-mass end ($\sigma \lesssim 0.5$, $M_h \gtrsim 10^{15}\,h^{-1}M_\odot$), where the super-exponential cutoff acts, is sparsely sampled: taking realisation 50 as an example, the most massive bins contain only $N = 65$, 6 and 2 haloes.
Evidence for the HMF cutoff also derives from concordance across the simulation suite: the same super-exponential structure is recovered independently across all 100 \texttt{Quijote} realisations.
This convergence to super-exponential cutoffs across all datasets analysed suggests that this functional structure is a robust empirical feature of the data.

\subsection{Comparison to alternative data}
\label{sec:alt_data}

Our results are derived from specific datasets---the \citet{Bernardi} SDSS photometry for the LF and SMF, and the \texttt{Quijote} HMF catalogues---so it is important to consider how they might change with alternative data.

\subsubsection{HMF}

The \texttt{Quijote} simulations have a particle mass of $8.207 \times 10^{10}\,h^{-1}M_\odot$, so the resolution limit of $\sim\!50$ particles per halo restricts us to $\sigma \lesssim 2$ and the extrapolation behaviour to larger $\sigma$ (lower masses) is not data-driven.
As shown in Fig.~\ref{fig:extrap}, the top four ESR functions all cut off sharply at large $\sigma$, with $f(\sigma) \to 0$, due to their super-exponential structure (whether of the form $\sim\!|\theta_0|^{g(\sigma)}$ or $\sim\!e^{-g(\sigma)}$).

Zoom simulations challenge this extrapolation. \citet{Wang2020} simulated haloes spanning 20 orders of magnitude down to Earth-mass scales ($\sim\!10^{-6}\,M_\odot$) using the Voids-within-Voids-within-Voids technique, and \citet{Zheng2024} used the same simulations to test the Press--Schechter, Sheth--Tormen, and extended Press--Schechter
formulae across this full mass range. They find that the extended Press--Schechter formula reproduces the simulation results to $\lesssim\!20\%$ accuracy from $10^{-6}$ to $10^{12.5}\,M_\odot$, supporting the continued $1/\sigma$ decline of $f(\sigma)$ at large $\sigma$ implied by the Press--Schechter $f \sim \delta_c/\sigma$ scaling (equivalently, a power-law rise in $n(M_h)$ as halo mass decreases). This is in tension with the ESR cutoff: if $f(\sigma)$ truly continues to rise at $\sigma \gg 2$, then the ESR functions---while optimal within the \texttt{Quijote} data range---would underpredict the abundance of low-mass haloes in the extrapolation regime.
This highlights an important limitation of data-driven function discovery: ESR identifies the functional forms that best compress the available data in an information-theoretic sense, but cannot be expected to predict behaviour in regimes far beyond that data unless that behaviour is captured by the simplicity criterion inherent in the DL metric. While ESR produces functions with Press--Schechter-like large-$\sigma$ scaling (e.g. Eq.~\ref{eq:hmf_ps_like}), they rank below those with a cutoff. If the mass range were extended to lower masses, ESR would likely promote such functions or identify new forms that accommodate the continued rise of $f(\sigma)$ at large $\sigma$.

\subsubsection{LF and SMF}

The LF and SMF have similar faint-end extrapolation concerns. The SDSS data we use are complete down to $M_\star \sim 10^{8.5}\,M_\odot$ and $L \sim 10^{8.5}\,L_\odot$. Deeper surveys such as GAMA \citep{Baldry2012, Wright2017} have measured the SMF down to $M_\star \sim 10^{7.5}\,M_\odot$, finding a steepening faint-end slope ($\approx -1.5$ logarithmically) in disagreement with the flat extrapolations of ESR (Fig.~\ref{fig:extrap}) and generally taken to suggest a double Schechter function.
If our ESR functions were fitted to such deeper data, the additional low-mass bins would promote or enforce such behaviour leading to different preferred functional forms.
Indeed a key advantage of ESR is that it can be readily applied to any such data, allowing an exploration not only of datasets' or mass ranges' relative parameter preferences within a given functional form, but their \emph{functional} preferences too.

An additional concern for the LF and SMF is the impact of the photometric and stellar population modelling choices involved. \citet{Bernardi} showed that the massive/bright end of both the LF and SMF is sensitive to the model used to fit galaxies' surface brightness profiles: single-S\'ersic fits recover significantly more light than the standard SDSS pipeline cmodel photometry, boosting $\phi$ at the bright end and flattening the exponential cutoff. Subsequent work has reinforced this finding. \citet{Fischer2017} showed that SDSS sky-subtraction biases the photometry of large galaxies, and that the choice of profile model (de~Vaucouleurs, S\'ersic, or S\'ersic-Exponential) significantly affects the recovered luminosity at the bright end. \citet{Bernardi2017b} demonstrated that these photometric differences propagate into the galaxy clustering signal and are not resolved by passive evolution corrections, implying that the systematic is astrophysically significant rather than a mere fitting artefact. \citet{Kravtsov2018} found that pipeline photometry underestimates the stellar masses of brightest cluster galaxies by factors of 2--4, because standard apertures miss the extended low-surface-brightness envelopes that S\'ersic fits capture.
Our explicit investigation of both S\'ersic and cmodel photometry goes some way towards characterising this systematic, but is by no means comprehensive.

For the SMF specifically, the conversion from luminosity to stellar mass introduces additional uncertainty through the assumed stellar population synthesis (SPS) model and initial mass function (IMF). \citet{Conroy2009} showed that SPS uncertainties alone can shift stellar mass estimates by $\sim\!0.3$\,dex, which would modify the shape of the SMF at both ends. Our analysis assumes the mass-to-light ratios provided by \citet{Bernardi} and does not vary or marginalise over SPS model uncertainty; doing so could in principle shift the preferred ESR functional form or parameter values.

\subsection{Additional caveats and systematics}
\label{sec:caveats}

Once the data and its error model are specified, ESR requires only two inputs from the user: the operator set from which to build the functions and the maximum complexity searched. Both of these can affect the results. Functions requiring operators outside the basis set (e.g.\ error functions, incomplete gamma functions) cannot be discovered, and thus better functions containing such operators may exist. The operator set also determines the complexities of functions: a function that is simple in one basis (e.g. a hyperbolic function in a basis containing such an operator) may be complex in another (requiring the hyperbolic function to be written out as exponentials). Regarding the maximum complexity (10), Fig.~\ref{fig:Pareto} shows that the DL is clearly flattening out at the high-complexity end for each dataset, even reaching a turnover at complexity 9 for LF S\'ersic. This suggests that we may be close to finding the true minimum description length function, beyond which the increase in complexity outweighs the increase in accuracy and the DL degrades. However, this is unlikely to be the case for the SMF: the fact that the Bernardi function gives significantly lower DL than any ESR function suggests the presence of many low-DL functions beyond complexity 10.

The DL metric we use (Eq.~\ref{eq:mdl}) effectively imposes a prior on functions given by~\citep{Priors}
\begin{equation}
    -\ln(P(f_i)) = k \ln(n) + \sum_j c_j,
\end{equation}
penalising functions purely on the basis of structural complexity. This allows, and can even favour, functions with bizarre structures unlikely to arise from any physical theory---an example is the nested exponential in the rank-2 function for LF cmodel (Table~\ref{tab:LF_functions}). An alternative is to use a language model such as the Katz back-off model to assign priors to functions on the basis of the frequency of their operator combinations in a training set of functions that have previously been proposed or found to be useful in the domain in question~\citep{Priors}. This upweights more ``reasonable'' functions~\citep{Priors,Sousa_2023,Martin_1}. We note, however, that our physicality checks perform a somewhat analogous role in selecting desirable functions beyond the requirement of low DL. The rank-2 LF cmodel function, for example, has a divergent integral, leading us not to emphasise it despite its accurate and structurally simple fit to the data.

It is important to note that the relative weight of the likelihood and complexity terms in the DL depends on the absolute galaxy/halo counts through the size of the Poisson uncertainties. At fixed LF/SMF/HMF shape, fewer objects and hence larger uncertainties reduce the magnitude of the likelihood, causing the complexity penalty to become relatively more important and hence simpler functions becoming favoured. Technically our results are therefore specific to the SDSS sample size and the \texttt{Quijote} box volume, although the features that likely determine the best-fit functions in reality (the functions' shapes and the fact that uncertainties grow at larger $L/M_\star/M_h$) are not dataset-dependent.

Our analysis assumes pre-calculated forms for the LF and SMF, including a prior accounting for selection effects and that the bins are entirely independent with Poisson-distributed counts. A more principled and accurate approach would be to forward-model the trial functions all the way to the data space and compare directly against individual galaxies' redshifts and fluxes. This would use the LF or SMF as the prior on galaxies' latent luminosities or stellar masses, and would allow principled selection modelling at the object level~\citep{Kelly_2008}. This is less of an issue for the HMF as there are no selection effects in simulated data, although mode-coupling through nonlinear gravitational evolution could still cause separated bins to covary so that the likelihood does not strictly factorise into a product over bins.

To assess the effect of this, we use the 100 \texttt{Quijote} realisations to compute the empirical covariance matrix of the halo counts across the 15 sigma bins common to all 100 sims. The diagonal elements (variances) are consistent with Poisson expectations: the variance-to-mean ratio ranges from 0.72 to 1.46 across bins, with a median of 0.95. The off-diagonal elements are non-negligible: the median absolute correlation coefficient is $|\rho_{ij}| = 0.12$, with a maximum of $|\rho| \approx 0.34$ between low-mass bins at $\sigma \approx 1.25$--$1.77$. 58\% of bin pairs have $|\rho| > 0.1$, and none exceed $0.5$. We then compare the ranking of the top ESR and literature HMF functions under both the Poisson likelihood and a Gaussian likelihood incorporating the full covariance matrix. The top-5 ESR functions are preserved under both metrics, with the best function $\theta_0/(\theta_1 + e^{\sigma^{\theta_2+\sigma}})$ remaining the top-ranked by NLL. Minor reorderings occur only among lower-ranked functions well below the top 5. We therefore conclude that the Poisson likelihood is adequate for the purpose of function selection, although the bin correlations would need to be accounted for in any application requiring precise parameter uncertainties or absolute goodness-of-fit assessment. There are also bin-to-bin covariances in the LF and SMF due to uncertainties in light profile shapes and mass-to-light ratios that correlate the bins; these cannot be estimated directly from the data, but are similarly expected not to be significant.

Finally, the parametric complexity term in Eq.~\ref{eq:mdl} uses only the diagonal elements $\hat{I}_{ii}$ of the Fisher information matrix rather than the full determinant $\det(\hat{I})$. This diagonal approximation is standard in the MDL literature~\citep{RISSANEN1978, MDL_review2}, but is strictly suitable only when
the parameters are uncorrelated at the maximum-likelihood point. When parameters are correlated, however, $\prod_i \hat{I}_{ii} > \det(\hat{I})$, so the diagonal approximation overestimates the parametric complexity and hence the DL. The full result is presented in the context of ESR in~\citet{Priors}.

To assess the effect of this we recompute the full Hessian matrix for the top 10 ESR functions and all literature functions on each dataset, and compare the resulting codelength
$\frac{1}{2}\ln\det(\hat{I})$ with the diagonal approximation $\sum_i\frac{1}{2}\ln\hat{I}_{ii}$. For the 3--4-parameter ESR functions the correction is small: typically 1--7\,nats for the LF and SMF, and 1--4\,nats per simulation for the HMF. For Schechter the correction is $\sim\!0.7$\,nats. The largest corrections are for the 7-parameter Bernardi function ($\sim\!15$--17\,nats, reflecting its richer parameter correlations) and for the Warren HMF fit ($\sim\!7$\,nats per simulation, i.e.\ $\sim\!700$ across 100 realisations, owing to the $\theta_0$--$\theta_2$ degeneracy). Crucially, however, these corrections do not change the ranking of the top-5 functions for any dataset: the shifts are correlated between functions and typically smaller than the raw DL differences.
We therefore conclude that the diagonal Fisher approximation does not materially affect the function rankings we report.

\subsection{Future work}
\label{sec:future}

Several extensions of this work would be valuable:

\begin{itemize}
\item \textbf{HMF universality across cosmologies}: The HMF is known to depend on cosmological parameters, and there is debate over whether it is truly ``universal'' (i.e.\ independent of cosmology) when expressed as $f(\sigma)$ \citep{Jenkins_2001, Tinker_HMF, Despali2016}. Running ESR on HMF data from simulations with different cosmologies (e.g.\ the full \texttt{Quijote} Latin hypercube) could test whether the same functional forms are preferred across cosmologies or whether cosmology-dependent terms emerge. \citet{Li_Smith} provide a framework for assessing universality that could be adapted, although including cosmological parameters as independent variables in the symbolic regression would require a different algorithm such as Operon~\citep{Operon}.

\item \textbf{Extended fitting range}: The extrapolation behaviour of the top ESR functions, while typically not leading to pathologies, is likely driven by the preference for simplicity encoded in the DL metric rather than the data. As discussed in Sec.~\ref{sec:alt_data}, data at lower $L$, $M_\star$ or $M_h$ tend to show power-law behaviour, captured by some of the literature fits but not by most of the ESR functions. Extending the data range would build this behaviour into the likelihood, improving the extrapolation of ESR functions.

\item \textbf{Redshift evolution}: The LF, SMF and HMF all evolve with redshift. Running ESR on data at multiple redshifts could characterise that evolution.

\item \textbf{Different wavelengths and surveys}: The LF depends on the photometric band \citep{Blanton_2003, Driver2012}, while the SMF depends on the assumed IMF and mass-to-light prescription. Running ESR on UV, infrared, or H$\alpha$ LFs from surveys such as GALEX \citep{Martin2005}, WISE \citep{Wright2010}, or upcoming facilities such as \emph{Euclid} \citep{Euclid2024} and the Vera C.\ Rubin Observatory's Legacy Survey of Space and Time \citep[LSST;][]{Ivezic2019}---and on SMFs from a range of stellar mass models---would test the generality of our results. Symbolic regression can tell us whether these variations, as well as those discussed above, are functional (i.e. the optimal functional form changes) or merely parametric (only the optimal parameter values vary within a fixed best-fit functional form).

\item \textbf{Extending to higher complexity}: For computational reasons ESR is restricted to complexity $\approx10$, which is not typically sufficient to identify a clear minimum in the DL-vs-complexity plot indicative of the true minimum-description-length function (Fig.~\ref{fig:Pareto}). Upcoming improvements to the efficiency of the ESR function generation and parameter fitting will allow it to reach complexity $\approx13$ (Kronberger et al., in prep). To reach higher complexities, genetic programming could be used (e.g. Operon), although this is not exhaustive and hence may fail to find some good functions.

\end{itemize}

\section{Conclusion}
\label{sec:conc}

Astrophysics is replete with fitting functions for datasets and processes that derive primarily from visual inspection. The optimality of such functions is rarely quantified, or uncertainties in the functional form propagated through analyses. This can be addressed by \emph{symbolic regression}, which quantifies the aptitude of a great range of functions automatically.

We use the ESR algorithm to optimise functional fits to the LF, SMF and HMF, using for the former S\'ersic and cmodel fits to SDSS $r$-band photometry~\citep{Bernardi} and for the latter the \texttt{Quijote} suite of $N$-body simulations~\citep{Quijote}. ESR operates only on relatively low-complexity functions for computational reasons, here extending to complexity 10. Functions are ranked by means of \emph{description length} (DL), describing the number of nats of information needed to express the data with the help of the function. This balances in a principled way functions' accuracies (the maximum log-likelihood of the data across the function's parameter space) and complexities, both structural and parametric. An independent set of ``physicality checks'' assesses extrapolation behaviour to weed out functions that fit the data well and are simple, but do not generalise sensibly beyond the data range. We similarly evaluate standard literature fitting functions---the Schechter~\citep{Schechter}, double Schechter and Bernardi~\citep{Bernardi} forms for the LF and SMF, and Press--Schechter~\citep{Press-Schechter}, Tinker~\citep{Tinker_HMF} and Warren~\citep{Warren_HMF} forms for the HMF.

For the LF, SMF and HMF, ESR discovers many functional forms with lower DL than the literature fits: the DLs and log-likelihoods are shown on a Pareto plot in Fig.~\ref{fig:Pareto}, and the best functions are overlaid on the data in Figs.~\ref{fig:LF_SMF}--\ref{fig:HMF_functions}. Many of these functions additionally pass the physicality checks, making them well-behaved alternatives to the literature functions that are superior in accuracy and complexity.
We caution however that the faint-end extrapolation (typically flat in $L$ and $M_\star$ and cut off in $M_h$) is driven by the preference for simplicity and is unlikely to represent galaxies and haloes accurately in that regime. The SMF appears to have additional structure that favours significantly more complex functions, so that the more complex form proposed by~\citet{Bernardi} performs better than any function found by ESR.

The ESR functions that we identify as the best overall are given in Eqs.~\ref{eq:1} (LF), \ref{eq:good} (SMF/cmodel LF), \ref{eq:2} (HMF) and \ref{eq:hmf_ps_like} (PS-like HMF); we recommend these as alternatives to the standard literature forms. Across all datasets these typically have a super-exponential cutoff at the bright/high-mass end (i.e.\ faster than the simple exponential of the Schechter function), driven by a stretched-exponential term $\sim\!|\theta|^{x^{\theta'}}$ with $\theta' \approx 0.3$--$0.5$.
We further assess consistency of the best ESR functions against all 100 \texttt{Quijote} boxes (differing only in random initial conditions), finding minimal differences such that the function ranking is largely robust across them (Fig.~\ref{fig:HMF_comp}): the rank-1 function is ranked top in 98 of 100 realisations, and 3$^\text{rd}$ and 4$^\text{th}$ in the others.

Our results lay the groundwork for using symbolic regression---and ESR specifically---to upgrade fitting functions found throughout astrophysics. This principled and automatic approach affords the optimal solution to fitting function problems, and enables uncertainties in functional forms as well as parameter values to be correctly propagated through larger analysis pipelines. This has strong implications for the accuracy of astrophysical studies.

\section*{Data availability}

The LF and SMF data are available as supplementary material in~\citet{Bernardi}, and the \texttt{Quijote} simulation data used to make the HMFs are available at \url{https://quijote-simulations.readthedocs.io}. ESR is available at \url{https://github.com/DeaglanBartlett/ESR}. The code used for this project is publicly available at \url{https://github.com/harrydesmond/LF_SMF_HMF_ESR}.

\section*{Acknowledgements}

We thank Richard Stiskalek for providing the \texttt{Quijote} halo mass functions.

AF was supported by a SEPNet summer research internship at the Institute of Cosmology and Gravitation, University of Portsmouth. HD is supported by a Royal Society University Research Fellowship (grant no. 211046). DJB was supported by the Simons Collaboration on ``Learning the Universe'' and acknowledges that support was provided by Schmidt Sciences, LLC. PGF acknowledges support from STFC and the Beecroft Trust.

\appendix
\section{HMF results extending to lower-mass haloes}
\label{app:hmf_full}

\begin{figure}
\includegraphics[width=\columnwidth]{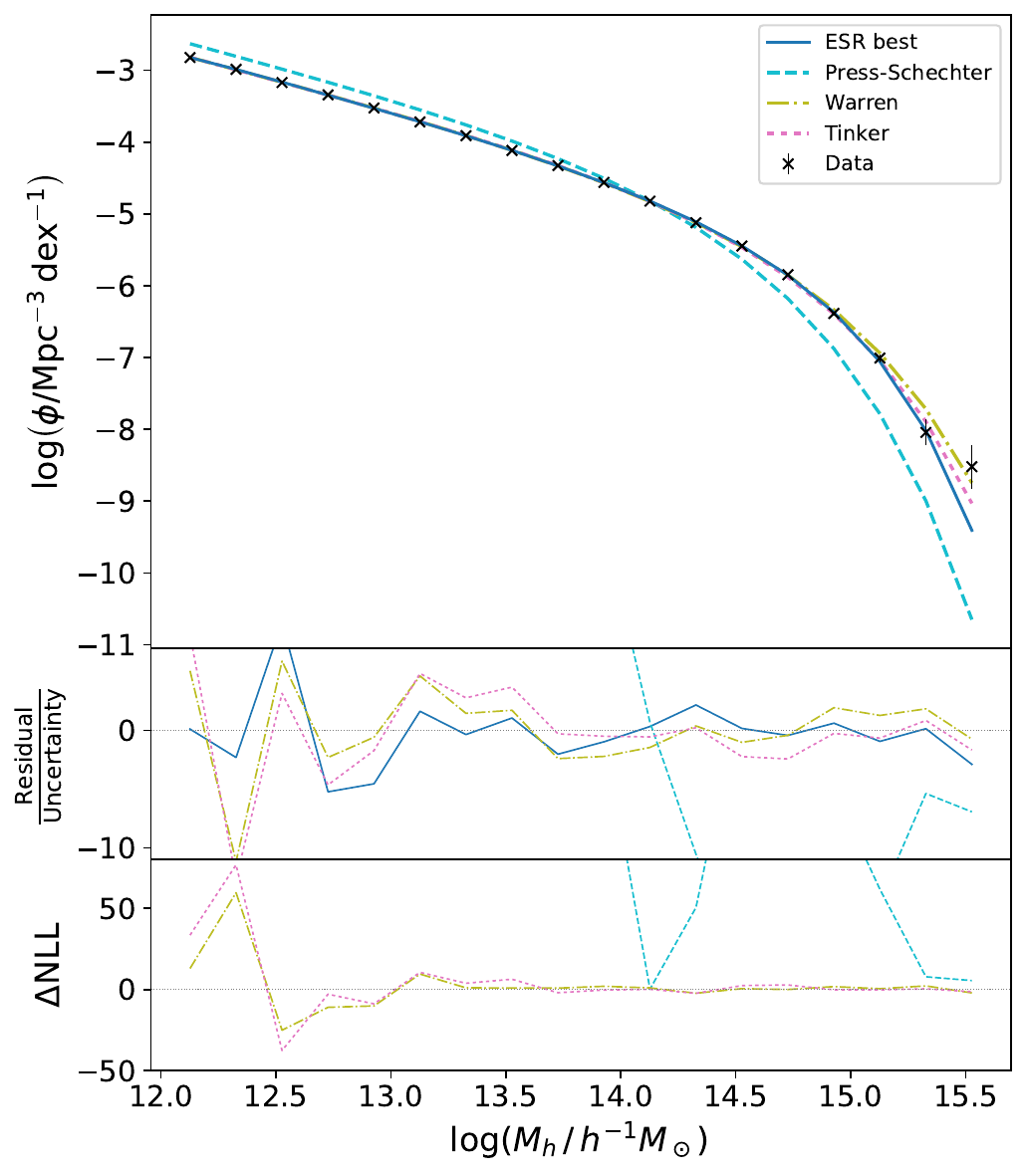}
\caption{As Fig.~\ref{fig:HMF_functions} but for the extended HMF dataset including two further low-mass bins. Note that the ``ESR best'' function shown here is different to that in Fig.~\ref{fig:HMF_functions} (top rows of Tables~\ref{tab:HMF_functions_full} and~\ref{tab:HMF_functions} respectively).}
\label{fig:HMF_functions_full}
\end{figure}

\begin{figure}
\includegraphics[width=\columnwidth]{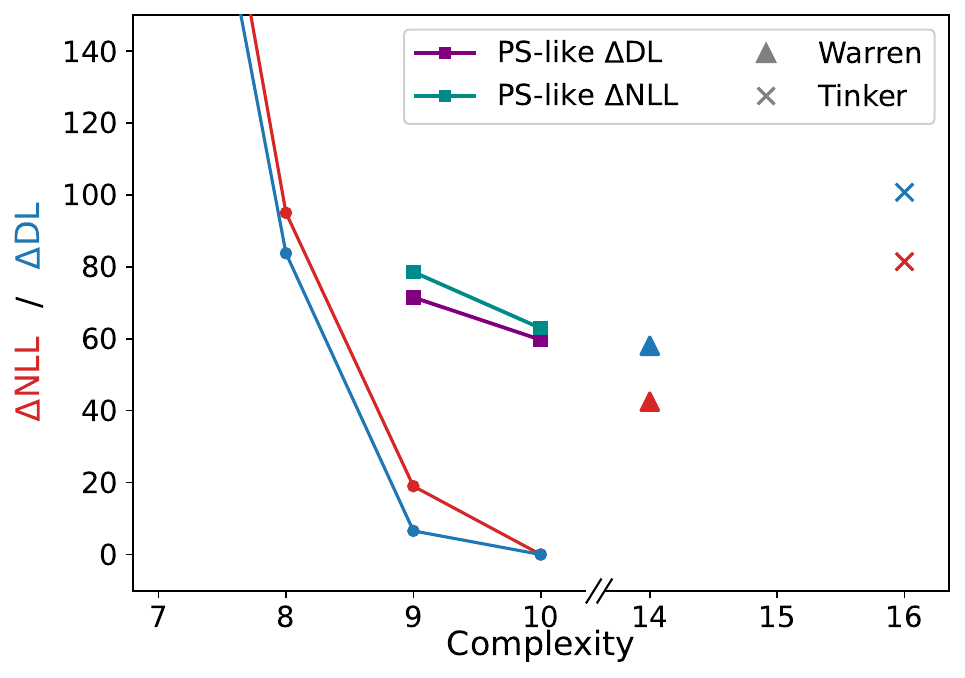}
\caption{As the HMF panel of Fig.~\ref{fig:Pareto} but for the extended mass range. Press--Schechter again lies far above the plotted range ($\Delta\mathrm{DL} \approx 2.96 \times 10^7$).
}
\label{fig:Pareto_full}
\end{figure}

\begin{figure*}
\includegraphics[width=\textwidth]{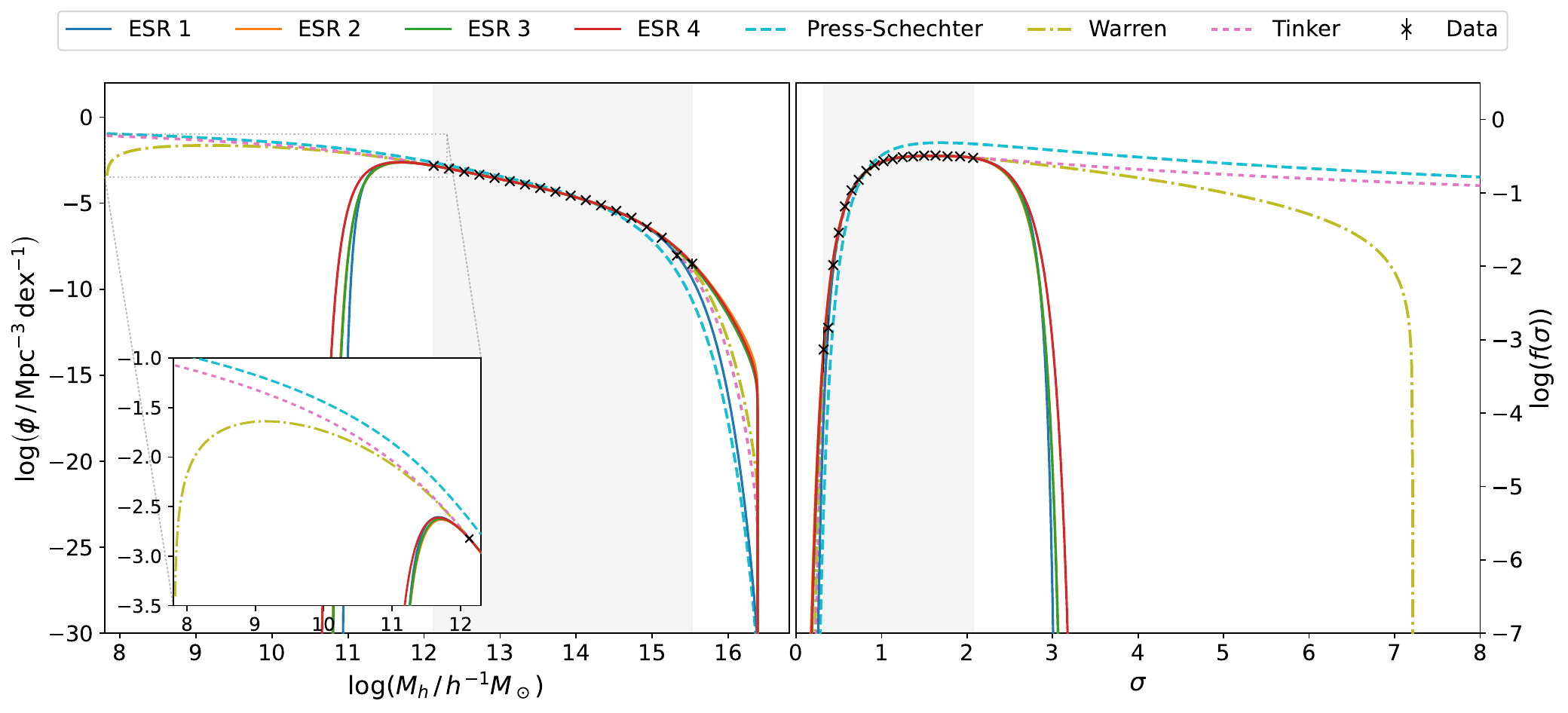}
\caption{As the HMF panels of Fig.~\ref{fig:extrap} but using the extended mass range.}
\label{fig:extrap_full}
\end{figure*}

\begin{figure*}
\includegraphics[width=1.\linewidth]{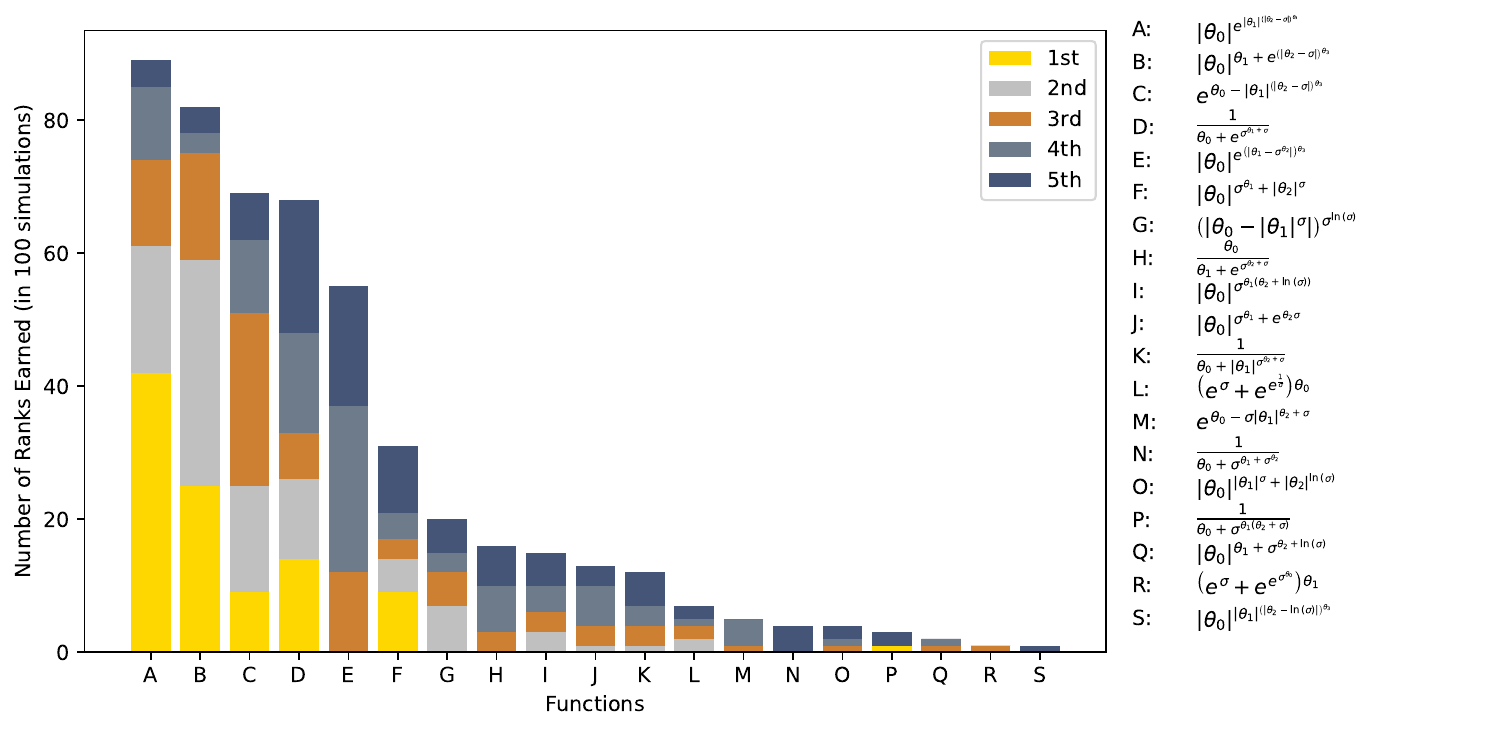}
\caption{As Fig.~\ref{fig:HMF_comp} but for the extended mass range.}
\label{fig:HMF_comp_full}
\end{figure*}

The HMF results in the main text use haloes down to $\sim50$ particles. Here we present the corresponding results with a further two 0.2 dex mass bins, going down to $\log(M_h/h^{-1}M_\odot) = 12.2$ or $\sim\!19$ particles. This provides a sensitivity test of our results, while revealing the kind of biases one may get when including poorly-resolved haloes. Compare Table~\ref{tab:app} to Table~\ref{tab:HMF_functions}, Fig.~\ref{fig:HMF_functions_full} to Fig.~\ref{fig:HMF_functions}, Fig.~\ref{fig:Pareto_full} to Fig.~\ref{fig:Pareto}, Fig.~\ref{fig:extrap_full} to Fig.~\ref{fig:extrap} and Fig.~\ref{fig:HMF_comp_full} to Fig.~\ref{fig:HMF_comp}.

The key differences are:
\begin{enumerate}
	\item The function rankings are substantially reshuffled, with a rank correlation of only $\rho \approx 0.65$ between the fiducial and extended rankings. The extended-range rank-1 function $|\theta_0|^{\exp(|\theta_1|^{|\theta_2 - \sigma|^{\theta_3}})}$ is not among the top 200 functions retained by the fiducial ESR search, and the fiducial rank-1 function $\theta_0/(\theta_1 + e^{\sigma^{\theta_2 + \sigma}})$ (3 parameters) is rank 13 on the extended data.
	\item The ESR advantage over the literature fits is larger over the extended mass range ($\Delta\mathrm{DL} \approx 4300$--$8400$ vs.\ $2200$--$3100$ for the fiducial data). This suggests that over the extended range ESR overfits the poorly-resolved haloes (which are very numerous, leading to very small Poisson uncertainties), while the literature functions have been chosen not to be so sensitive. This can be clearly seen in the spike in $\Delta$NLL in the two lowest-mass bins in Fig.~\ref{fig:HMF_functions_full} (note the change of y-scale relative to Fig.~\ref{fig:HMF_functions}). In Fig.~\ref{fig:Pareto_full}, the PS-like functions lie further above the ESR Pareto front, indicating that this (physical) behaviour is not preferred by the two lowest mass bins.
	\item The consistency across realisations is weaker on the extended dataset (e.g. the top-ranked function appears 89/100 times in the top five and 42/100 times at rank 1, versus 100/100 and 98/100 respectively over the fiducial range). This indicates a degree of inconsistency of those lowest two mass bins with the rest of the relation, likely due to insufficient resolution, causing an inter-realisation scatter in the function fits.
	\item The collapsed mass fractions are higher on the extended dataset ($\sim\!0.4$--$0.7$ vs.\ $0.36$--$0.47$), indicating that the lowest two mass bins are pushing the fitted HMFs up.
\end{enumerate}

\begin{table*}
\centering
\caption{Top ESR functions for the HMF over the extended mass range (cf. Table~\ref{tab:HMF_functions}).}
\label{tab:HMF_functions_full}

\begin{tabular}{>{\rowstyle}c>{\rowstyle}l>{\rowstyle}c>{\rowstyle}c>{\rowstyle}c>{\rowstyle}c>{\rowstyle}c>{\rowstyle}c>{\rowstyle}c>{\rowstyle}c}
\hline
Rank & Function & Comp. & $\theta_0$ & $\theta_1$ & $\theta_2$ & $\theta_3$ & $\Delta$DL & $\Delta$NLL & Checks \\
\hline \hline
1 & $|\theta_0|^{\exp\left(|\theta_1|^{|\theta_2 - \sigma|^{\theta_3}}\right)}$ & 10 & $0.66^{+0.00}_{-0.00}$ & $1.70^{+0.01}_{-0.01}$ & $1.63^{+0.00}_{-0.00}$ & $2.78^{+0.02}_{-0.02}$ & 0 & 0 & \checkmark \\
2 & $e^{\theta_0 - |\theta_1|^{|\theta_2 - \sigma|^{\theta_3}}}$ & 10 & $-0.15^{+0.00}_{-0.00}$ & $2.27^{+0.02}_{-0.02}$ & $1.62^{+0.00}_{-0.00}$ & $3.27^{+0.02}_{-0.03}$ & 191 & 922 & \checkmark \\
3 & $|\theta_0|^{\theta_1 + e^{|\theta_2 - \sigma|^{\theta_3}}}$ & 10 & $0.49^{+0.01}_{-0.01}$ & $0.60^{+0.03}_{-0.03}$ & $1.63^{+0.00}_{-0.00}$ & $3.07^{+0.03}_{-0.03}$ & 332 & 137 & \checkmark \\
4 & $|\theta_0|^{\exp\left(|\theta_1 - \sigma^{\theta_2}|^{\theta_3}\right)}$ & 10 & $0.32^{+0.00}_{-0.00}$ & $1.56^{+0.00}_{-0.00}$ & $0.92^{+0.00}_{-0.00}$ & $3.21^{+0.03}_{-0.03}$ & 878 & 900 & \checkmark \\
5 & $|\theta_0 - |\theta_1|^\sigma|^{\sigma^{\ln\sigma}}$ & 7 & $0.94^{+0.00}_{-0.00}$ & $0.68^{+0.00}_{-0.00}$ & --- & --- & 1404 & 2277 & \checkmark \\
6 & $|\theta_0|^{|\theta_1|^{|\theta_2 - \ln\sigma|^{\theta_3}}}$ & 10 & $0.32^{+0.00}_{-0.00}$ & $2.33^{+0.02}_{-0.02}$ & $0.46^{+0.00}_{-0.00}$ & $2.05^{+0.01}_{-0.01}$ & 1419 & 1692 & \checkmark \\
7 & $\frac{1}{\theta_0 + e^{\sigma^{\sigma - \theta_1}}}$ & 9 & $1.10^{+0.00}_{-0.00}$ & $2.37^{+0.00}_{-0.00}$ & --- & --- & 1502 & 2555 & \checkmark \\
8 & $e^{\theta_0 - (\sigma|\theta_1|)^{\theta_2 + \sigma}}$ & 10 & $-0.71^{+0.01}_{-0.01}$ & $1.28^{+0.01}_{-0.01}$ & $-2.78^{+0.01}_{-0.01}$ & --- & 2078 & 2242 & \checkmark \\
\hline
$>200$ & P.~Sch. (Eq.~\ref{eq:f_ps}) & 10 & $\delta_c = 1.686$ & --- & --- & --- & $2.96 \times 10^7$ & $2.96 \times 10^7$ & \checkmark \\
\noalign{\gmark}
--- & Warren (Eq.~\ref{eq:f_warren}) & 14 & $7.43^{+0.70}_{-0.61}$ & $-0.92^{+0.01}_{-0.01}$ & $-0.04^{+0.00}_{-0.00}$ & $0.80^{+0.00}_{-0.01}$ & 4343 & 3389 & (a) \\
--- & Tinker (Eq.~\ref{eq:f_tinker}) & 16 & $8.9^{+2.9}_{-2.7} \times 10^{-4}$ & $0.80^{+0.01}_{-0.01}$ & $5300^{+2200}_{-2100}$ & $0.93^{+0.01}_{-0.01}$ & 8358 & 7433 & (b) \\
\hline \noalign{\nmark}
\end{tabular}

\smallskip
\noindent \emph{Checks:} \checkmark\, = all physicality checks passed: $f(\sigma) \to 0$ as $\sigma \to \infty$, $f(\sigma)$ remains finite as $\sigma \to 0^+$, and $\int f(\sigma)\,\mathrm{d}\ln\sigma$ converges to a value $\leq 1$ (the collapsed mass fraction).
(a)~$f(\sigma) < 0$ for $\sigma \gtrsim 7$;
$\int f\,\mathrm{d}\ln\sigma$ is negative.
(b)~$f \to \theta_0 > 0$ as $\sigma \to \infty$, so $f$ does not strictly vanish; however $\int f\,\mathrm{d}\ln\sigma$ converges ($\approx 0.79$).
\label{tab:app}
\end{table*}

\bibliographystyle{mnras}
\bibliography{references}

\begin{thebibliography}{}
\makeatletter
\relax
\def\mn@urlcharsother{\let\do\@makeother \do\$\do\&\do\#\do\^\do\_\do\%\do\~}
\def\mn@doi{\begingroup\mn@urlcharsother \@ifnextchar [ {\mn@doi@}
  {\mn@doi@[]}}
\def\mn@doi@[#1]#2{\def\@tempa{#1}\ifx\@tempa\@empty \href
  {http://dx.doi.org/#2} {doi:#2}\else \href {http://dx.doi.org/#2} {#1}\fi
  \endgroup}
\def\mn@eprint#1#2{\mn@eprint@#1:#2::\@nil}
\def\mn@eprint@arXiv#1{\href {http://arxiv.org/abs/#1} {{\tt arXiv:#1}}}
\def\mn@eprint@dblp#1{\href {http://dblp.uni-trier.de/rec/bibtex/#1.xml}
  {dblp:#1}}
\def\mn@eprint@#1:#2:#3:#4\@nil{\def\@tempa {#1}\def\@tempb {#2}\def\@tempc
  {#3}\ifx \@tempc \@empty \let \@tempc \@tempb \let \@tempb \@tempa \fi \ifx
  \@tempb \@empty \def\@tempb {arXiv}\fi \@ifundefined
  {mn@eprint@\@tempb}{\@tempb:\@tempc}{\expandafter \expandafter \csname
  mn@eprint@\@tempb\endcsname \expandafter{\@tempc}}}

\bibitem[\protect\citeauthoryear{{Allen}, {Evrard}  \& {Mantz}}{{Allen}
  et~al.}{2011}]{Allen2011}
{Allen} S.~W.,  {Evrard} A.~E.,   {Mantz} A.~B.,  2011, \mn@doi [\araa]
  {10.1146/annurev-astro-081710-102514}, 49, 409

\bibitem[\protect\citeauthoryear{{Baldry} et~al.,}{{Baldry}
  et~al.}{2012}]{Baldry2012}
{Baldry} I.~K.,  et~al., 2012, \mn@doi [\mnras]
  {10.1111/j.1365-2966.2012.20340.x}, 421, 621

\bibitem[\protect\citeauthoryear{Bartlett, Desmond  \& Ferreira}{Bartlett
  et~al.}{2022}]{ESR}
Bartlett D.~J.,  Desmond H.,   Ferreira P.~G.,  2022, \mn@doi [IEEE
  Transactions on Evolutionary Computation] {10.1109/TEVC.2023.3280250}

\bibitem[\protect\citeauthoryear{Bartlett, Desmond  \& Ferreira}{Bartlett
  et~al.}{2023a}]{ESR_priors}
Bartlett D.~J.,  Desmond H.,   Ferreira P.~G.,  2023a, in {The Genetic and
  Evolutionary Computation Conference 2023}.  (\mn@eprint {arXiv}
  {2304.06333}), \mn@doi{10.1145/3583133.3596327}

\bibitem[\protect\citeauthoryear{{Bartlett}, {Desmond}  \&
  {Ferreira}}{{Bartlett} et~al.}{2023b}]{Priors}
{Bartlett} D.~J.,  {Desmond} H.,   {Ferreira} P.~G.,  2023b, \mn@doi [arXiv
  e-prints] {10.48550/arXiv.2304.06333}, \href
  {https://ui.adsabs.harvard.edu/abs/2023arXiv230406333B} {p. arXiv:2304.06333}

\bibitem[\protect\citeauthoryear{{Behroozi}, {Wechsler}  \&
  {Conroy}}{{Behroozi} et~al.}{2013a}]{Behroozi2013}
{Behroozi} P.~S.,  {Wechsler} R.~H.,   {Conroy} C.,  2013a, \mn@doi [\apj]
  {10.1088/0004-637X/770/1/57}, 770, 57

\bibitem[\protect\citeauthoryear{{Behroozi}, {Wechsler}  \&
  {Conroy}}{{Behroozi} et~al.}{2013b}]{Behroozi_2013}
{Behroozi} P.~S.,  {Wechsler} R.~H.,   {Conroy} C.,  2013b, \mn@doi [\apj]
  {10.1088/0004-637X/770/1/57}, 770, 57

\bibitem[\protect\citeauthoryear{{Bell}, {McIntosh}, {Katz}  \&
  {Weinberg}}{{Bell} et~al.}{2003}]{Bell_2003}
{Bell} E.~F.,  {McIntosh} D.~H.,  {Katz} N.,   {Weinberg} M.~D.,  2003, \mn@doi
  [\apjs] {10.1086/378847}, 149, 289

\bibitem[\protect\citeauthoryear{{Benson}, {Bower}, {Frenk}, {Lacey}, {Baugh}
  \& {Cole}}{{Benson} et~al.}{2003}]{Benson2003}
{Benson} A.~J.,  {Bower} R.~G.,  {Frenk} C.~S.,  {Lacey} C.~G.,  {Baugh} C.~M.,
    {Cole} S.,  2003, \mn@doi [\apj] {10.1086/379160}, 599, 38

\bibitem[\protect\citeauthoryear{{Bernardi}, {Meert}, {Sheth}, {Vikram},
  {Huertas-Company}, {Mei}  \& {Shankar}}{{Bernardi} et~al.}{2013}]{Bernardi}
{Bernardi} M.,  {Meert} A.,  {Sheth} R.~K.,  {Vikram} V.,  {Huertas-Company}
  M.,  {Mei} S.,   {Shankar} F.,  2013, \mn@doi [\mnras]
  {10.1093/mnras/stt1607}, \href
  {https://ui.adsabs.harvard.edu/abs/2013MNRAS.436..697B} {436, 697}

\bibitem[\protect\citeauthoryear{{Bernardi}, {Fischer}, {Sheth}, {Meert},
  {Huertas-Company}, {Shankar}  \& {Vikram}}{{Bernardi}
  et~al.}{2017}]{Bernardi2017b}
{Bernardi} M.,  {Fischer} J.-L.,  {Sheth} R.~K.,  {Meert} A.,
  {Huertas-Company} M.,  {Shankar} F.,   {Vikram} V.,  2017, \mn@doi [\mnras]
  {10.1093/mnras/stx275}, 468, 2569

\bibitem[\protect\citeauthoryear{Biggio, Bendinelli, Neitz, Lucchi  \&
  Parascandolo}{Biggio et~al.}{2021}]{Biggio_2021}
Biggio L.,  Bendinelli T.,  Neitz A.,  Lucchi A.,   Parascandolo G.,  2021, in
  Meila M.,  Zhang T.,  eds,  Proceedings of Machine Learning Research Vol.
  139, Proceedings of the 38th International Conference on Machine Learning.
  PMLR, pp 936--945, \url {https://proceedings.mlr.press/v139/biggio21a.html}

\bibitem[\protect\citeauthoryear{{Blanton} \& {Roweis}}{{Blanton} \&
  {Roweis}}{2007}]{Blanton_Roweis_2007}
{Blanton} M.~R.,  {Roweis} S.,  2007, \mn@doi [\aj] {10.1086/510127}, 133, 734

\bibitem[\protect\citeauthoryear{{Blanton} et~al.,}{{Blanton}
  et~al.}{2003}]{Blanton_2003}
{Blanton} M.~R.,  et~al., 2003, \mn@doi [\apj] {10.1086/375776}, \href
  {https://ui.adsabs.harvard.edu/abs/2003ApJ...592..819B} {592, 819}

\bibitem[\protect\citeauthoryear{{Blanton}, {Kazin}, {Muna}, {Weaver}  \&
  {Price-Whelan}}{{Blanton} et~al.}{2011}]{Blanton2011}
{Blanton} M.~R.,  {Kazin} E.,  {Muna} D.,  {Weaver} B.~A.,   {Price-Whelan} A.,
   2011, \mn@doi [\aj] {10.1088/0004-6256/142/1/31}, 142, 31

\bibitem[\protect\citeauthoryear{{Bower}, {Benson}, {Malbon}, {Helly}, {Frenk},
  {Baugh}, {Cole}  \& {Lacey}}{{Bower} et~al.}{2006}]{Bower_2006}
{Bower} R.~G.,  {Benson} A.~J.,  {Malbon} R.,  {Helly} J.~C.,  {Frenk} C.~S.,
  {Baugh} C.~M.,  {Cole} S.,   {Lacey} C.~G.,  2006, \mn@doi [\mnras]
  {10.1111/j.1365-2966.2006.10519.x}, 370, 645

\bibitem[\protect\citeauthoryear{Broyden}{Broyden}{1970}]{BFGS_1}
Broyden C.~G.,  1970, \mn@doi [IMA Journal of Applied Mathematics]
  {10.1093/imamat/6.1.76}, 6, 76

\bibitem[\protect\citeauthoryear{Burlacu, Kronberger  \& Kommenda}{Burlacu
  et~al.}{2020}]{Operon}
Burlacu B.,  Kronberger G.,   Kommenda M.,  2020, in Proceedings of the 2020
  Genetic and Evolutionary Computation Conference Companion. GECCO '20.
Association for Computing Machinery, New York, NY, USA, p. 1562–1570,
  \mn@doi{10.1145/3377929.3398099}, \url
  {https://doi.org/10.1145/3377929.3398099}

\bibitem[\protect\citeauthoryear{{Conroy}, {Gunn}  \& {White}}{{Conroy}
  et~al.}{2009}]{Conroy2009}
{Conroy} C.,  {Gunn} J.~E.,   {White} M.,  2009, \mn@doi [\apj]
  {10.1088/0004-637X/699/1/486}, 699, 486

\bibitem[\protect\citeauthoryear{{Cooray} \& {Sheth}}{{Cooray} \&
  {Sheth}}{2002}]{2002PhR...372....1C}
{Cooray} A.,  {Sheth} R.,  2002, \mn@doi [\physrep]
  {10.1016/S0370-1573(02)00276-4}, \href
  {https://ui.adsabs.harvard.edu/abs/2002PhR...372....1C} {372, 1}

\bibitem[\protect\citeauthoryear{Cranmer}{Cranmer}{2020}]{pysr}
Cranmer M.,  2020, PySR: Fast \& Parallelized Symbolic Regression in
  Python/Julia, \mn@doi{10.5281/zenodo.4041459}, \url
  {http://doi.org/10.5281/zenodo.4041459}

\bibitem[\protect\citeauthoryear{{Croton} et~al.,}{{Croton}
  et~al.}{2006}]{Croton_2006}
{Croton} D.~J.,  et~al., 2006, \mn@doi [\mnras]
  {10.1111/j.1365-2966.2005.09675.x}, 365, 11

\bibitem[\protect\citeauthoryear{David}{David}{1989}]{David}
David E.,  1989, Genetic Algorithms in Search, Optimization and Machine
  Learning.
Addison-Wesley

\bibitem[\protect\citeauthoryear{{Davis}, {Efstathiou}, {Frenk}  \&
  {White}}{{Davis} et~al.}{1985}]{FOF}
{Davis} M.,  {Efstathiou} G.,  {Frenk} C.~S.,   {White} S.~D.~M.,  1985,
  \mn@doi [\apj] {10.1086/163168}, \href
  {https://ui.adsabs.harvard.edu/abs/1985ApJ...292..371D} {292, 371}

\bibitem[\protect\citeauthoryear{{Desmond}}{{Desmond}}{2025}]{Desmond_ESR}
{Desmond} H.,  2025, \mn@doi [arXiv e-prints] {10.48550/arXiv.2507.13033},
  \href {https://ui.adsabs.harvard.edu/abs/2025arXiv250713033D} {p.
  arXiv:2507.13033}

\bibitem[\protect\citeauthoryear{{Desmond}, {Bartlett}  \&
  {Ferreira}}{{Desmond} et~al.}{2023}]{ESR_on_RAR}
{Desmond} H.,  {Bartlett} D.~J.,   {Ferreira} P.~G.,  2023, arXiv e-prints,
  \href {https://ui.adsabs.harvard.edu/abs/2023arXiv230104368D} {p.
  arXiv:2301.04368}

\bibitem[\protect\citeauthoryear{{Despali}, {Giocoli}, {Angulo}, {Tormen},
  {Sheth}, {Baso}  \& {Moscardini}}{{Despali} et~al.}{2016}]{Despali2016}
{Despali} G.,  {Giocoli} C.,  {Angulo} R.~E.,  {Tormen} G.,  {Sheth} R.~K.,
  {Baso} G.,   {Moscardini} L.,  2016, \mn@doi [\mnras]
  {10.1093/mnras/stv2842}, 456, 2486

\bibitem[\protect\citeauthoryear{{Driver}, {Robotham}, {Kelvin}, {Alpaslan},
  {Baldry}  et~al.}{{Driver} et~al.}{2012}]{Driver2012}
{Driver} S.~P.,  {Robotham} A.~S.~G.,  {Kelvin} L.,  {Alpaslan} M.,  {Baldry}
  I.~K.,   et~al., 2012, \mn@doi [\mnras] {10.1111/j.1365-2966.2012.22036.x},
  427, 3244

\bibitem[\protect\citeauthoryear{{Eckert}, {Kannappan}, {Stark}, {Moffett},
  {Berlind}  \& {Norris}}{{Eckert} et~al.}{2016}]{Eckert_BMF}
{Eckert} K.~D.,  {Kannappan} S.~J.,  {Stark} D.~V.,  {Moffett} A.~J.,
  {Berlind} A.~A.,   {Norris} M.~A.,  2016, \mn@doi [\apj]
  {10.3847/0004-637X/824/2/124}, \href
  {https://ui.adsabs.harvard.edu/abs/2016ApJ...824..124E} {824, 124}

\bibitem[\protect\citeauthoryear{{Euclid Collaboration}, {Mellier}
  et~al.}{{Euclid Collaboration} et~al.}{2025}]{Euclid2024}
{Euclid Collaboration} {Mellier} Y.,   et~al., 2025, \mn@doi [\aap]
  {10.1051/0004-6361/202450810}, 697, A1

\bibitem[\protect\citeauthoryear{{Fischer}, {Bernardi}  \& {Meert}}{{Fischer}
  et~al.}{2017}]{Fischer2017}
{Fischer} J.-L.,  {Bernardi} M.,   {Meert} A.,  2017, \mn@doi [\mnras]
  {10.1093/mnras/stx138}, 467, 490

\bibitem[\protect\citeauthoryear{Fletcher}{Fletcher}{1970}]{BFGS_2}
Fletcher R.,  1970, \mn@doi [The Computer Journal] {10.1093/comjnl/13.3.317},
  13, 317

\bibitem[\protect\citeauthoryear{Grunwald}{Grunwald}{2007}]{MDL_review2}
Grunwald P.,  2007, The Minimum Description Length Principle.
MIT Press

\bibitem[\protect\citeauthoryear{{Gr{\"u}nwald} \& {Roos}}{{Gr{\"u}nwald} \&
  {Roos}}{2019}]{MDL_review1}
{Gr{\"u}nwald} P.,  {Roos} T.,  2019, arXiv e-prints, \href
  {https://ui.adsabs.harvard.edu/abs/2019arXiv190808484G} {p. arXiv:1908.08484}

\bibitem[\protect\citeauthoryear{Haupt \& Haupt}{Haupt \& Haupt}{2004}]{haupt}
Haupt R.,  Haupt S.,  2004, Practical genetic algorithms, 2nd edn.
Wyley

\bibitem[\protect\citeauthoryear{{Henriques}, {White}, {Thomas}, {Angulo},
  {Guo}, {Lemson}, {Springel}  \& {Overzier}}{{Henriques}
  et~al.}{2015}]{Henriques_2015}
{Henriques} B. M.~B.,  {White} S. D.~M.,  {Thomas} P.~A.,  {Angulo} R.,  {Guo}
  Q.,  {Lemson} G.,  {Springel} V.,   {Overzier} R.,  2015, \mn@doi [\mnras]
  {10.1093/mnras/stv705}, 451, 2663

\bibitem[\protect\citeauthoryear{{Ivezi{\'{c}}}, {Kahn}, {Tyson}
  et~al.}{{Ivezi{\'{c}}} et~al.}{2019}]{Ivezic2019}
{Ivezi{\'{c}}} {\v{Z}}.,  {Kahn} S.~M.,  {Tyson} J.~A.,   et~al., 2019, \mn@doi
  [\apj] {10.3847/1538-4357/ab042c}, 873, 111

\bibitem[\protect\citeauthoryear{{Jenkins}, {Frenk}, {White}, {Colberg},
  {Cole}, {Evrard}, {Couchman}  \& {Yoshida}}{{Jenkins}
  et~al.}{2001}]{Jenkins_2001}
{Jenkins} A.,  {Frenk} C.~S.,  {White} S.~D.~M.,  {Colberg} J.~M.,  {Cole} S.,
  {Evrard} A.~E.,  {Couchman} H.~M.~P.,   {Yoshida} N.,  2001, \mn@doi [\mnras]
  {10.1046/j.1365-8711.2001.04029.x}, 321, 372

\bibitem[\protect\citeauthoryear{{Jin}, {Fu}, {Kang}, {Guo}  \& {Guo}}{{Jin}
  et~al.}{2019}]{Jin_2019}
{Jin} Y.,  {Fu} W.,  {Kang} J.,  {Guo} J.,   {Guo} J.,  2019, \mn@doi [arXiv
  e-prints] {10.48550/arXiv.1910.08892}, \href
  {https://ui.adsabs.harvard.edu/abs/2019arXiv191008892J} {p. arXiv:1910.08892}

\bibitem[\protect\citeauthoryear{{Kelly}, {Fan}  \& {Vestergaard}}{{Kelly}
  et~al.}{2008}]{Kelly_2008}
{Kelly} B.~C.,  {Fan} X.,   {Vestergaard} M.,  2008, \mn@doi [\apj]
  {10.1086/589501}, \href
  {https://ui.adsabs.harvard.edu/abs/2008ApJ...682..874K} {682, 874}

\bibitem[\protect\citeauthoryear{{Kravtsov}, {Vikhlinin}  \&
  {Meshcheryakov}}{{Kravtsov} et~al.}{2018}]{Kravtsov2018}
{Kravtsov} A.~V.,  {Vikhlinin} A.~A.,   {Meshcheryakov} A.~V.,  2018, \mn@doi
  [Astronomy Letters] {10.1134/S1063773717120015}, 44, 8

\bibitem[\protect\citeauthoryear{{Kronberger}, {Burlacu}, {Kommenda}, {Winkler}
   \& {Affenzeller}}{{Kronberger} et~al.}{2024a}]{Kronberger2024book}
{Kronberger} G.,  {Burlacu} B.,  {Kommenda} M.,  {Winkler} S.~M.,
  {Affenzeller} M.,  2024a, Symbolic Regression.
Genetic and Evolutionary Computation, Springer,
  \mn@doi{10.1007/978-3-031-56957-0}

\bibitem[\protect\citeauthoryear{{Kronberger}, {Olivetti de Franca}, {Desmond},
  {Bartlett}  \& {Kammerer}}{{Kronberger} et~al.}{2024b}]{Kronberger}
{Kronberger} G.,  {Olivetti de Franca} F.,  {Desmond} H.,  {Bartlett} D.~J.,
  {Kammerer} L.,  2024b, \mn@doi [arXiv e-prints] {10.48550/arXiv.2404.17292},
  \href {https://ui.adsabs.harvard.edu/abs/2024arXiv240417292K} {p.
  arXiv:2404.17292}

\bibitem[\protect\citeauthoryear{Landajuela et~al.,}{Landajuela
  et~al.}{2022}]{Landajuela_2022}
Landajuela M.,  et~al., 2022, in Proceedings of the 36th International
  Conference on Neural Information Processing Systems. NIPS '22.
Curran Associates Inc., Red Hook, NY, USA

\bibitem[\protect\citeauthoryear{{Li} \& {Smith}}{{Li} \&
  {Smith}}{2024}]{Li_Smith}
{Li} Y.,  {Smith} R.~E.,  2024, \mn@doi [arXiv e-prints]
  {10.48550/arXiv.2411.18722}, \href
  {https://ui.adsabs.harvard.edu/abs/2024arXiv241118722L} {p. arXiv:2411.18722}

\bibitem[\protect\citeauthoryear{{Li}, {Lelli}, {McGaugh}, {Pawlowski}, {Zwaan}
   \& {Schombert}}{{Li} et~al.}{2019}]{Li_HMF}
{Li} P.,  {Lelli} F.,  {McGaugh} S.,  {Pawlowski} M.~S.,  {Zwaan} M.~A.,
  {Schombert} J.,  2019, \mn@doi [\apjl] {10.3847/2041-8213/ab53e6}, \href
  {https://ui.adsabs.harvard.edu/abs/2019ApJ...886L..11L} {886, L11}

\bibitem[\protect\citeauthoryear{{Madau} \& {Dickinson}}{{Madau} \&
  {Dickinson}}{2014}]{Madau_2014}
{Madau} P.,  {Dickinson} M.,  2014, \mn@doi [\araa]
  {10.1146/annurev-astro-081811-125615}, \href
  {https://ui.adsabs.harvard.edu/abs/2014ARA&A..52..415M} {52, 415}

\bibitem[\protect\citeauthoryear{{Martin} et~al.,}{{Martin}
  et~al.}{2005}]{Martin2005}
{Martin} D.~C.,  et~al., 2005, \mn@doi [\apjl] {10.1086/426387}, 619, L1

\bibitem[\protect\citeauthoryear{{Mart{\'\i}n}, {Yasin}, {Bartlett}, {Desmond}
  \& {Ferreira}}{{Mart{\'\i}n} et~al.}{2025}]{Martin_2}
{Mart{\'\i}n} A.,  {Yasin} T.,  {Bartlett} D.~J.,  {Desmond} H.,   {Ferreira}
  P.~G.,  2025, \mn@doi [arXiv e-prints] {10.48550/arXiv.2511.23073}, \href
  {https://ui.adsabs.harvard.edu/abs/2025arXiv251123073M} {p. arXiv:2511.23073}

\bibitem[\protect\citeauthoryear{{Mart{\'\i}n}, {Yasin}, {Bartlett}, {Desmond}
  \& {Ferreira}}{{Mart{\'\i}n} et~al.}{2026}]{Martin_1}
{Mart{\'\i}n} A.,  {Yasin} T.,  {Bartlett} D.~J.,  {Desmond} H.,   {Ferreira}
  P.~G.,  2026, \mn@doi [arXiv e-prints] {10.48550/arXiv.2601.05203}, \href
  {https://ui.adsabs.harvard.edu/abs/2026arXiv260105203M} {p. arXiv:2601.05203}

\bibitem[\protect\citeauthoryear{{Moster}, {Naab}  \& {White}}{{Moster}
  et~al.}{2013a}]{Moster2013}
{Moster} B.~P.,  {Naab} T.,   {White} S. D.~M.,  2013a, \mn@doi [\mnras]
  {10.1093/mnras/sts261}, 428, 3121

\bibitem[\protect\citeauthoryear{{Moster}, {Naab}  \& {White}}{{Moster}
  et~al.}{2013b}]{Moster_2013}
{Moster} B.~P.,  {Naab} T.,   {White} S. D.~M.,  2013b, \mn@doi [\mnras]
  {10.1093/mnras/sts261}, 428, 3121

\bibitem[\protect\citeauthoryear{{Murray}}{{Murray}}{2014}]{hmf}
{Murray} S.,  2014, {HMF: Halo Mass Function calculator}, Astrophysics Source
  Code Library, record ascl:1412.006 (\mn@eprint {ascl} {1412.006})

\bibitem[\protect\citeauthoryear{{Murray}, {Power}  \& {Robotham}}{{Murray}
  et~al.}{2013}]{HMFcalc}
{Murray} S.~G.,  {Power} C.,   {Robotham} A.~S.~G.,  2013, \mn@doi [Astronomy
  and Computing] {10.1016/j.ascom.2013.11.001}, \href
  {https://ui.adsabs.harvard.edu/abs/2013A&C.....3...23M} {3, 23}

\bibitem[\protect\citeauthoryear{{Papastergis}, {Cattaneo}, {Huang},
  {Giovanelli}  \& {Haynes}}{{Papastergis} et~al.}{2012}]{Papastergis_BMF}
{Papastergis} E.,  {Cattaneo} A.,  {Huang} S.,  {Giovanelli} R.,   {Haynes}
  M.~P.,  2012, \mn@doi [\apj] {10.1088/0004-637X/759/2/138}, \href
  {https://ui.adsabs.harvard.edu/abs/2012ApJ...759..138P} {759, 138}

\bibitem[\protect\citeauthoryear{{Peebles}}{{Peebles}}{1980}]{1980lssu.book.....P}
{Peebles} P.~J.~E.,  1980, {The large-scale structure of the universe}

\bibitem[\protect\citeauthoryear{Petersen, Larma, Mundhenk, Santiago, Kim  \&
  Kim}{Petersen et~al.}{2021}]{Petersen_2021}
Petersen B.~K.,  Larma M.~L.,  Mundhenk T.~N.,  Santiago C.~P.,  Kim S.~K.,
  Kim J.~T.,  2021, in International Conference on Learning Representations.
  \url {https://openreview.net/forum?id=m5Qsh0kBQG}

\bibitem[\protect\citeauthoryear{{Pillepich} et~al.,}{{Pillepich}
  et~al.}{2018}]{Pillepich_2018}
{Pillepich} A.,  et~al., 2018, \mn@doi [\mnras] {10.1093/mnras/stx2656}, 473,
  4077

\bibitem[\protect\citeauthoryear{{Planck Collaboration} et~al.,}{{Planck
  Collaboration} et~al.}{2020}]{Planck18_Parameters}
{Planck Collaboration} et~al., 2020, \mn@doi [\aap]
  {10.1051/0004-6361/201833910}, \href
  {https://ui.adsabs.harvard.edu/abs/2020A&A...641A...6P} {641, A6}

\bibitem[\protect\citeauthoryear{Ponomareva et~al.,}{Ponomareva
  et~al.}{2023}]{Anastasia_HIMF}
Ponomareva A.~A.,  et~al., 2023, \mn@doi [Monthly Notices of the Royal
  Astronomical Society] {10.1093/mnras/stad1249}, 522, 5308

\bibitem[\protect\citeauthoryear{{Press} \& {Schechter}}{{Press} \&
  {Schechter}}{1974}]{Press-Schechter}
{Press} W.~H.,  {Schechter} P.,  1974, \mn@doi [\apj] {10.1086/152650}, \href
  {https://ui.adsabs.harvard.edu/abs/1974ApJ...187..425P} {187, 425}

\bibitem[\protect\citeauthoryear{Rissanen}{Rissanen}{1978}]{RISSANEN1978}
Rissanen J.,  1978, \mn@doi [Automatica]
  {https://doi.org/10.1016/0005-1098(78)90005-5}, 14, 465

\bibitem[\protect\citeauthoryear{{Rozo} et~al.,}{{Rozo}
  et~al.}{2010}]{Rozo2010}
{Rozo} E.,  et~al., 2010, \mn@doi [\apj] {10.1088/0004-637X/708/1/645}, 708,
  645

\bibitem[\protect\citeauthoryear{{Schaye} et~al.,}{{Schaye}
  et~al.}{2015}]{Schaye_2015}
{Schaye} J.,  et~al., 2015, \mn@doi [\mnras] {10.1093/mnras/stu2058}, 446, 521

\bibitem[\protect\citeauthoryear{{Schechter}}{{Schechter}}{1976}]{Schechter}
{Schechter} P.,  1976, \mn@doi [\apj] {10.1086/154079}, \href
  {https://ui.adsabs.harvard.edu/abs/1976ApJ...203..297S} {203, 297}

\bibitem[\protect\citeauthoryear{{Schmidt}}{{Schmidt}}{1968}]{Schmidt_1968}
{Schmidt} M.,  1968, \mn@doi [\apj] {10.1086/149446}, 151, 393

\bibitem[\protect\citeauthoryear{{Sheth} \& {Tormen}}{{Sheth} \&
  {Tormen}}{1999}]{Sheth_Tormen_1999}
{Sheth} R.~K.,  {Tormen} G.,  1999, \mn@doi [\mnras]
  {10.1046/j.1365-8711.1999.02692.x}, 308, 119

\bibitem[\protect\citeauthoryear{{Sousa}, {Bartlett}, {Desmond}  \&
  {Ferreira}}{{Sousa} et~al.}{2023}]{Sousa_2023}
{Sousa} T.,  {Bartlett} D.~J.,  {Desmond} H.,   {Ferreira} P.~G.,  2023,
  \mn@doi [arXiv e-prints] {10.48550/arXiv.2310.16786}, \href
  {https://ui.adsabs.harvard.edu/abs/2023arXiv231016786S} {p. arXiv:2310.16786}

\bibitem[\protect\citeauthoryear{{Springel} et~al.,}{{Springel}
  et~al.}{2008}]{Springel}
{Springel} V.,  et~al., 2008, \mn@doi [\mnras]
  {10.1111/j.1365-2966.2008.14066.x}, \href
  {https://ui.adsabs.harvard.edu/abs/2008MNRAS.391.1685S} {391, 1685}

\bibitem[\protect\citeauthoryear{{Tenachi}, {Ibata}  \&
  {Diakogiannis}}{{Tenachi} et~al.}{2023}]{Tenachi_2023}
{Tenachi} W.,  {Ibata} R.,   {Diakogiannis} F.~I.,  2023, \mn@doi [arXiv
  e-prints] {10.48550/arXiv.2303.03192}, \href
  {https://ui.adsabs.harvard.edu/abs/2023arXiv230303192T} {p. arXiv:2303.03192}

\bibitem[\protect\citeauthoryear{{Tinker}, {Kravtsov}, {Klypin}, {Abazajian},
  {Warren}, {Yepes}, {Gottl{\"o}ber}  \& {Holz}}{{Tinker}
  et~al.}{2008}]{Tinker_HMF}
{Tinker} J.,  {Kravtsov} A.~V.,  {Klypin} A.,  {Abazajian} K.,  {Warren} M.,
  {Yepes} G.,  {Gottl{\"o}ber} S.,   {Holz} D.~E.,  2008, \mn@doi [\apj]
  {10.1086/591439}, \href
  {https://ui.adsabs.harvard.edu/abs/2008ApJ...688..709T} {688, 709}

\bibitem[\protect\citeauthoryear{Turing}{Turing}{1950}]{turing}
Turing A.~M.,  1950, \mn@doi [Mind] {10.1093/mind/LIX.236.433}, LIX, 433

\bibitem[\protect\citeauthoryear{{Vikhlinin} et~al.,}{{Vikhlinin}
  et~al.}{2009}]{Vikhlinin_2009}
{Vikhlinin} A.,  et~al., 2009, \mn@doi [\apj] {10.1088/0004-637X/692/2/1060},
  692, 1060

\bibitem[\protect\citeauthoryear{{Villaescusa-Navarro}
  et~al.,}{{Villaescusa-Navarro} et~al.}{2020}]{Quijote}
{Villaescusa-Navarro} F.,  et~al., 2020, \mn@doi [\apjs]
  {10.3847/1538-4365/ab9d82}, \href
  {https://ui.adsabs.harvard.edu/abs/2020ApJS..250....2V} {250, 2}

\bibitem[\protect\citeauthoryear{{Wang}, {Bose}, {Frenk}, {Gao}, {Jenkins},
  {Springel}  \& {White}}{{Wang} et~al.}{2020}]{Wang2020}
{Wang} J.,  {Bose} S.,  {Frenk} C.~S.,  {Gao} L.,  {Jenkins} A.,  {Springel}
  V.,   {White} S. D.~M.,  2020, \mn@doi [Nature] {10.1038/s41586-020-2642-9},
  585, 39

\bibitem[\protect\citeauthoryear{{Warren}, {Abazajian}, {Holz}  \&
  {Teodoro}}{{Warren} et~al.}{2006}]{Warren_HMF}
{Warren} M.~S.,  {Abazajian} K.,  {Holz} D.~E.,   {Teodoro} L.,  2006, \mn@doi
  [\apj] {10.1086/504962}, \href
  {https://ui.adsabs.harvard.edu/abs/2006ApJ...646..881W} {646, 881}

\bibitem[\protect\citeauthoryear{{Wechsler} \& {Tinker}}{{Wechsler} \&
  {Tinker}}{2018}]{Wechsler_Tinker_2018}
{Wechsler} R.~H.,  {Tinker} J.~L.,  2018, \mn@doi [\araa]
  {10.1146/annurev-astro-081817-051756}, 56, 435

\bibitem[\protect\citeauthoryear{{Wright} et~al.,}{{Wright}
  et~al.}{2010}]{Wright2010}
{Wright} E.~L.,  et~al., 2010, \mn@doi [\aj] {10.1088/0004-6256/140/6/1868},
  140, 1868

\bibitem[\protect\citeauthoryear{{Wright} et~al.,}{{Wright}
  et~al.}{2017}]{Wright2017}
{Wright} A.~H.,  et~al., 2017, \mn@doi [\mnras] {10.1093/mnras/stx1149}, 470,
  283

\bibitem[\protect\citeauthoryear{{York}, {Adelman}, {Anderson}, {Anderson},
  {Annis}, {Bahcall}  et~al.}{{York} et~al.}{2000}]{York2000}
{York} D.~G.,  {Adelman} J.,  {Anderson} Jr. J.~E.,  {Anderson} S.~F.,  {Annis}
  J.,  {Bahcall} N.~A.,   et~al., 2000, \mn@doi [\aj] {10.1086/301513}, 120,
  1579

\bibitem[\protect\citeauthoryear{{Zheng}, {Bose}, {Frenk}, {Gao}, {Jenkins},
  {Liao}, {Liu}  \& {Wang}}{{Zheng} et~al.}{2024}]{Zheng2024}
{Zheng} H.,  {Bose} S.,  {Frenk} C.~S.,  {Gao} L.,  {Jenkins} A.,  {Liao} S.,
  {Liu} Y.,   {Wang} J.,  2024, \mn@doi [\mnras] {10.1093/mnras/stae414}, 528,
  7300

\bibitem[\protect\citeauthoryear{{Zwaan}, {Meyer}, {Staveley-Smith}  \&
  {Webster}}{{Zwaan} et~al.}{2005}]{Zwaan_2005}
{Zwaan} M.~A.,  {Meyer} M.~J.,  {Staveley-Smith} L.,   {Webster} R.~L.,  2005,
  \mn@doi [\mnras] {10.1111/j.1745-3933.2005.00029.x}, 359, L30

\bibitem[\protect\citeauthoryear{{von der Linden} et~al.,}{{von der Linden}
  et~al.}{2014}]{vonderLinden2014}
{von der Linden} A.,  et~al., 2014, \mn@doi [\mnras] {10.1093/mnras/stt1945},
  439, 2

\makeatother
\end{thebibliography}

\label{lastpage}
\end{document}